\RequirePackage[2020-02-02]{latexrelease}
\documentclass[aps,twocolumn,showpacs,preprintnumbers,prb,superscriptaddress]{revtex4-2}

\usepackage{graphicx}  
\usepackage{multirow}
\usepackage{xcolor}

\usepackage{fancyhdr}
\usepackage{longtable}
\usepackage{parskip}
\usepackage[T1]{fontenc}
\usepackage{dcolumn}   

\usepackage{bm}        
\usepackage{amsfonts}  
\usepackage{amsmath}   
\usepackage{amssymb}   

\usepackage{natbib}

\newcommand{\pwisein}{\left\{ \begin{array}{ll}}
\newcommand{\pwiseout}{\end{array}\right.}

\begin{document}

\title{Exploring interacting chiral spin chains in terms of black hole physics}

\author{Ewan Forbes} 
\affiliation {\it School of Physics and Astronomy, University of Leeds, Leeds, LS2 9JT, United Kingdom}

\author{Matthew D. Horner}
\affiliation {\it School of Physics and Astronomy, University of Leeds, Leeds, LS2 9JT, United Kingdom}
\affiliation{Aegiq Ltd., Cooper Buildings, Arundel Street, Sheffield, S1 2NS, United Kingdom}

\author{Andrew Hallam} 
\affiliation {\it School of Physics and Astronomy, University of Leeds, Leeds, LS2 9JT, United Kingdom}

\author{Joseph Barker} 
\affiliation {\it School of Physics and Astronomy, University of Leeds, Leeds, LS2 9JT, United Kingdom}

\author{Jiannis K. Pachos}
\affiliation {\it School of Physics and Astronomy, University of Leeds, Leeds, LS2 9JT, United Kingdom}

\date{\today}

\begin{abstract}  

In this paper we explore the properties of a 1-dimensional spin chain in the presence of chiral interactions, focusing on the system's transition to distinct chiral phases for various values of the chiral coupling. By employing the mean field theory approximation we establish a connection between this chiral system and a Dirac particle in the curved spacetime of a black hole. Surprisingly, the black hole horizon coincides with the interface between distinct chiral phases. We examine the chiral properties of the system for homogeneous couplings and in scenarios involving position dependent couplings that correspond to black hole geometries. To determine the significance of interactions in the chiral chain we employ bosonization techniques and derive the corresponding Luttinger liquid model. Furthermore, we investigate the classical version of the model to understand the impact of the chiral operator on the spins and gain insight into the observed chirality. Our findings shed light on the behavior of the spin chain under the influence of the chiral operator, elucidating the implications of chirality in various contexts, including black hole physics.

\end{abstract}


\maketitle 

\section{Introduction}

An intriguing family of lattice models can be described by relativistic physics in their continuum limit. One prominent illustration of this phenomenon is graphene, whose behavior at low energy can be effectively described by the renowned Dirac equation \cite{Graphite,Graphene}. Similar relativistic descriptions can be found in diverse examples such as Kitaev's honeycomb model \cite{Kitaev_geo,KITAEV20062}, superconductors \cite{FQHE,Supercon}, and the XX model \cite{LIEB1961407,De_Pasquale_2008}. These relativistic frameworks not only deepen our understanding of these systems but also pave the way for the simulation of high-energy physics with table-top experiments.

In this paper, we explore a chiral modification of the 1D spin-1/2 XX model \cite{Mathew}. The XX model can be expressed in terms of free fermions and thus it is analytically tractable and well understood. The introduction of a three-spin chiral term renders it interacting and thus hard to investigate analytically or numerically. It is noteworthy that such chiral systems exhibit a rich spectrum of quantum correlations \cite{Entanglement} and they can give rise to skyrmions \cite{Tsomokos_2008}. Remarkably, we demonstrate that these chiral systems can be effectively modeled by the Dirac equation on a curved spacetime. This intriguing connection offers a unique opportunity to realize a black hole background within the laboratory setting.

The emergent black hole physics is explicitly revealed by applying the mean field (MF) approximation, and its existence can be verified by investigating the Hawking effect. Hawking radiation, resulting from vacuum fluctuations of quantum fields near a black hole's horizon, leads to the evaporation of the black hole \cite{Hawking:1975vcx,Page_2005}. The mechanism used to find this \cite{Mathew}, involving a wavepacket tunneling across the horizon and escaping with a thermal distribution, originally derived in \cite{Wilczek}, allows the simulation of Hawking radiation in fermionic lattice models \cite{Sabsovich,Maertens,BH_semimetal,Volovik1,Volovik2,Volovik3,Hang,Guan_2017,Retzker,Roldan-Molina,Rodriguez,Thermal,Stone_2013,Steinhauer_2016,Kosior,Yang}.
We test the reliability of this
approximation through a detailed analysis of the bosonization of the full spin model.
We find that the MF approximation faithfully predicts a
phase transition between a chiral and non-chiral phase.
Remarkably, the emergent event horizon aligns
with the interface between chiral and non-chiral phases. In particular, we find that the inside of the black hole
corresponds to a chiral region with a central charge of
$c = 2$ where the chiral interaction is dominant. The outside corresponds to a non-chiral region where the XX model is dominant,
with a central charge $c = 1$. Subsequently, we examine the MF approximation's validity by employing bosonization techniques, of which a general case has been studied for two \cite{miranda_2003,Giamarchi} and four \cite{four_fermi} Fermi points, which allow us to map the fully interacting Hamiltonian onto a Luttinger liquid \cite{Giamarchi}. Additionally, we employ a classical version of the system to further analyze and understand these effects in a comprehensive manner. By doing so, we gain valuable insights into the impact of chirality on the spins along the chain and its consequential effects on the entire system. We envision that the presented geometric description provides an elegant formalism
to model strongly interacting systems and their interfaces also in higher dimensions and thus predict their behaviour.

This article is organised as follows: In Section II we introduce our model, then diagonalize it to highlight the characteristics of the system, such as the transition in the dispersion relation due to the effect of the chiral order parameter. In Section III we give an effective description of the chiral chain in terms of black hole geometry. In Section IV we investigate the chiral spin operator and its expectation with the ground state of the system both in the flat and curved space cases. In Section V we use bosonization to ascertain the significance of the interactions, and in Section VI we use a classical version of the system to gain a geometric intuition on chiral interactions. We give concluding remarks and an outlook in Section VII.

\section{Chiral Chain Model}

Here we introduce the chiral spin chain, transform it to interacting fermions and then apply mean field theory to determine its properties.

\subsection{The mean field approximation}

\begin{figure}[t]
\begin{center}
\includegraphics[width=0.75\linewidth]{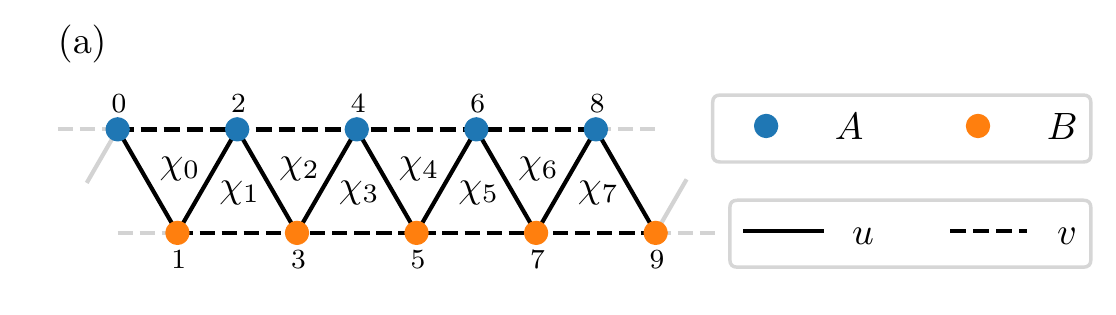}
\includegraphics[width=\linewidth]{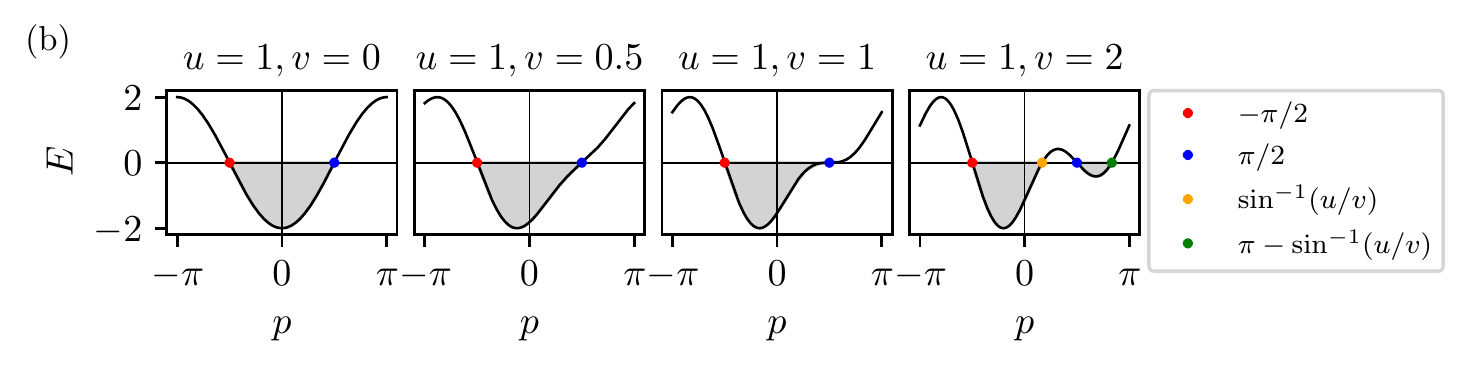}
\end{center}
\caption{(a) The interactions of the lattice diagrammatically portrayed with nearest neighbour interaction strength defined by $u$ and next-to-nearest neighbour interactions by $v$ \cite{Mathew}, with the seperations of neighbouring spins into groups $A$ and $B$ representing the unit cell described in Eq.~(\ref{eq:lattice_ham}). The chirality operator calculates the interaction of the three spins in each triangular space \cite{Tsomokos_2008}. (b) The dispersion relation of the Hamiltonian for various values of $v$. We see that two additional Fermi points appear if $v > u$ which divides the negative-energy portion of the Brillouin zone into two disconnected regions.}
\label{fig:C4_lattice}
\end{figure}

The system we investigate here is the one-dimensional spin-$\frac{1}{2}$ chain with the Hamiltonian
\begin{equation}
H =  \sum_{n=0}^{N-1} \left[- \frac{u}{2} \left(\sigma^x_n \sigma^x_{n+1} + \sigma^y_n \sigma^y_{n+1} \right) - \frac{v}{4} \chi_n \right], \label{eq:C4_spin_ham}
\end{equation}
where the spin chirality operator is \cite{Chiral_phase, Tsomokos_2008}
 \begin{equation}
\chi_n  = \vec{\sigma}_n \cdot \left(\vec{\sigma}_{n+1} \times \vec{\sigma}_{n+2}\right),
\label{eq:C4_chiral_op}
\end{equation}
where $\vec{\sigma}_n = (\sigma^x_n, \sigma^y_n, \sigma^z_n)$ is the spin vector of Pauli operators, and the $u,v$ couplings are real numbers with dimensions of energy. This model is a modified XX model with an additional three-spin interaction term $\chi$, as shown in Fig.~\ref{fig:C4_lattice}(a). Here, we adopt periodic boundary conditions with $\vec{\sigma}_N = \vec{\sigma}_0$. If we introducing $\sigma_n^\pm = (\sigma^x_n \pm i \sigma^y_n)/2$, then by employing the Jordan-Wigner transformation defined by $\sigma^+_n = (-1)^{\Sigma_n} c_n$ where $\Sigma_n =  \sum_{m < n} c_m^\dagger c_m$ and $\sigma^z = 2c^\dagger_n c_n - 1$ \cite{coleman_2015}, we can map the Hamiltonian to
\begin{equation}
\begin{split}
H  =  \sum_{n=0}^{N-1} \bigg\{ & - u c_n^\dagger c_{n+1} - \frac{iv}{2} c^\dagger_n c_{n+2}  \\
& - \frac{iv}{2}  \Big[ c_n^\dagger c_{n+1} (2c^\dagger_{n+2}c_{n+2} - 1) \\ & - c^\dagger_{n+1} c_{n+2} (2c^\dagger_nc_n - 1) \Big] \bigg\} + \text{H.c.}, \label{eq:C4_app_h_interacting} 
\end{split}
\end{equation} 
where  $c_n$ are a set of fermionic modes obeying the anti-commutation relations $\{ c_n , c_m \} = \{ c_m^\dagger , c_n^\dagger \} = 0$ and $\{ c_n , c_m^\dagger \} = \delta_{mn}$. We see that the model is intrinsically interacting as the fermionic Hamiltonian contains quartic terms.

To analyse the behaviour of the interacting model, we apply mean field theory (MFT) to transform the Hamiltonian into an effective quadratic Hamiltonian which can be efficiently diagonalised. MFT defines the fluctuation of an operator $A$ as $\delta A = A - \langle A \rangle $, where $\langle A \rangle$ is the expectation value of the operator $A$ with respect to the mean field ground state $|\Omega\rangle$. For a product of two operators we have
\begin{equation}
AB   = \langle A \rangle B + A \langle B \rangle - \langle A \rangle \langle B \rangle + \delta A \delta B,
\end{equation}
where the second order in fluctuations can be ignored. Applying this to the interacting terms of Eq.~(\ref{eq:C4_app_h_interacting}), the Hamiltonian becomes
\begin{equation}
\begin{split}
H_\mathrm{MF}(\alpha,Z) &= \sum_{n=0}^{N-1} \left[ -(u - iv Z) c^\dagger_n c_{n+1} - \frac{iv}{2} c^\dagger_n c_{n+2} \right] \\
&+ \mu \sum_{n=0}^{N-1} c^\dagger_n c_n + E_0 + \mathrm{H.c.}, \label{eq:H_MF_parametrised}
\end{split}
\end{equation}
where $\mu =  2 v \mathrm{Im}(\alpha)$ is an effective chemical potential controlling the number of particles in the ground state, $E_0 =  v ( Z - 1) \mathrm{Im}(\alpha) $ is a constant energy shift, and  $\langle \sigma_n^z \rangle = Z$, $\langle c^\dagger_n c_{n+1} \rangle = \alpha$, where the expectation value is done with respect to the ground state of the mean field Hamiltonian, $|\Omega(\alpha,Z) \rangle$, for given values of $\alpha$ and $Z$. Self consistency requires $\langle \Omega(\alpha,Z)| \sigma^z_n | \Omega(\alpha,Z) \rangle = Z $ and $\langle \Omega(\alpha,Z) | c^\dagger_n c_{n+1} |\Omega(\alpha,Z) \rangle  = \alpha$ for all $n$. While these two equations have many solutions, we can single one out on physical grounds: the original Hamiltonian of Eq.~(\ref{eq:C4_app_h_interacting}) has particle-hole symmetry, $[H,U]= 0$, where $U$ is the particle-hole transformation with $U c_n U^\dagger = (-1)^n c_n^\dagger$ and $U c_n^\dagger U^\dagger = (-1)^n c_n$. This symmetry implies that $\langle c_n^\dagger c_{n}\rangle = 1/2$ and $\langle c_n^\dagger c_{n+1} \rangle \in \mathbb{R}$ in the ground state. If we require the MFT to retain the particle-hole symmetry, then these conditions imply that $Z = \mu = \alpha = 0$, and the MFT Hamiltonian becomes
\begin{equation}
H_\text{MF} = \sum_{n = 0}^{N-1} \left( -u c^\dagger_n c_{n+1} - \frac{iv}{2} c^\dagger_n c_{n+2} \right) + \text{H.c.}. \label{eq:mf_ham}
\end{equation}
Another way to show the vanishing of $\alpha$ is that particle-hole symmetry implies that $\alpha$ must be real. However, in Eq.~(\ref{eq:H_MF_parametrised}) we see that $\alpha$ only appears via Im($\alpha$), which is zero if particle-hole symmetry is applied and therefore vanishes from $H$ fully.

This Hamiltonian is quadratic and periodic, hence it can be diagonalised with a Fourier transform
\begin{equation}
c_n = \frac{1}{\sqrt{N}} \sum_{p \in \mathrm{B.Z.}} e^{ianp} c_p, \label{eq:C4_fourier_transform}
\end{equation}
where $\mathrm{B.Z.} = [-\pi/a, \pi/a)$ is the Brillouin zone, $p$ are the momenta quantised as $p = 2n \pi/Na$ for $n \in \mathbb{Z}$, $c_p$ are the momentum space fermionic modes, and $a$ is the lattice spacing. This brings the Hamiltonian into the diagonal form 
\begin{equation}
H_\mathrm{MF} = \sum_{p \in \mathrm{B.Z.}} E(p) c_p^\dagger c_p,
\end{equation}
where the dispersion relation is given by
\begin{equation}
E(p) = -2u \cos(ap) + v \sin(2ap), \label{eq:C4_mono_dispersion}
\end{equation}
as shown in Fig.~\ref{fig:C4_lattice}. The Fermi points of this model, defined as the points $\{ p_i \}$ such that $E(p_i) = 0$, are given  by $p_\pm = \pm \pi/2a$ for $|v| < |u|$, whilst for $|v| \geq |u|$ we find two additional Fermi points located at
\begin{equation}
p_1 = \frac{1}{a} \sin^{-1} \left( \frac{u}{v} \right) , \quad p_2 = \frac{\pi}{a} - p_1 ,
\end{equation}
as shown in Fig.~\ref{fig:C4_lattice}(b). These additional points arise due to the Nielsen-Ninomiya theorem.

\subsection{Phase transitions \label{section:phase_transitions}}

To investigate the nature of quantum phases supported by Eq.~(\ref{eq:C4_spin_ham}), and the transitions between them, we consider the case of homogeneous couplings $u$ and $v$ along the chain. In this section, we focus on the predictions of the mean field Hamiltonian of Eq.~(\ref{eq:mf_ham}) and compare it with the results obtained using matrix product state analysis of the spin Hamiltonian of Eq.~(\ref{eq:C4_spin_ham}) \cite{Mathew}. All analytic calculations of this section are done using the mean field theory.
\subsubsection{Correlations}
The correlation matrix is defined as $C_{nm} = \langle \Omega | c^\dagger_n c_m |\Omega \rangle$, where $|\Omega \rangle$ is the ground state of Hamiltonian (\ref{eq:mf_ham}). Mapping to momentum space with a discrete Fourier transform as in Eq.~(\ref{eq:C4_fourier_transform}), we can write
\begin{equation}
\begin{aligned}
C_{nm} & = \frac{1}{N} \sum_{p,q \in \text{B.Z.}} e^{-i p n} e^{i q m} \langle \Omega | c^\dagger_p c_q |\Omega \rangle \\
&  = \frac{1}{2 \pi} \sum_{p:E(p)<0} \Delta p e^{-ip(n-m)}  \\
& = \frac{1}{2 \pi} \int_{p:E(p)< 0} \mathrm{d}p e^{-ip(n-m)},
\end{aligned} \label{eq:C4_correlation_def}
\end{equation} 
where in the second equality we used the fact that the ground state $|\Omega\rangle$ has all negative energy states occupied, so $\langle \Omega | c^\dagger_p c_q |\Omega \rangle = \delta_{pq} \theta(-E(p))$ and used the fact that eigenstates are separated in momentum space by $\Delta p = 2 \pi/N$ for a lattice spacing $a = 1$ to rewrite the sum as a Riemann sum. In the third equality we took the thermodynamic limit $N\rightarrow \infty$ mapping the sum to an integral which can now be solved analytically.

For $|v| < |u|$ the correlation function is given by
\begin{equation}
C_{nm} = \frac{\sin\left[ \frac{\pi}{2} (n-m) \right] }{\pi (n-m)}. \label{eq:C4_correlation_v<u}
\end{equation}
For $|v| > |u|$, the negative energy portion of the Brillouin zone splits into two disconnected regions so the integral splits into two as
\begin{equation}
\begin{aligned}
C_{nm} & =\frac{1}{2\pi} \left( \int_{-\frac{\pi}{2}}^{p_1} \mathrm{d}p + \int_{\frac{\pi}{2}}^{\pi - p_1} \mathrm{d}p \right) e^{-ipa(n-m)} \\
& = \frac{i}{2 \pi (n-m)} \bigg\{ -2 \cos\left[ (n-m)\frac{\pi}{2} \right] \\
&+ (-1)^{n-m} e^{i p_1 (n-m)} + e^{-ip_1(n-m)}\bigg\}, \label{eq:C4_correlation_v>u}
\end{aligned} 
\end{equation}
which is now a function of $v$ and is not smooth. The fact the correlation matrix is not a smooth function of $v$ is a consequence of the change in topology of the Fermi sea shown by the grey portion of Fig.~\ref{fig:C4_lattice}. As observables are derived from the correlation matrix, this behaviour is the root cause of the phase transition exhibited by the model.

\subsubsection{Energy density}
 
The ground state $|\Omega\rangle$ is the state for which the Fermi sea is fully occupied. Therefore, the ground state energy density is given by
\begin{equation}
\rho_0  = \frac{1}{N} \langle \Omega | H|\Omega\rangle \rightarrow \frac{1}{2 \pi} \int_{p:E(p)<0} \mathrm{d}p E(p),
\end{equation}
where we took the thermodynamic limit $N \rightarrow \infty$ by using the standard trick of moulding the sum into a Riemann sum and taking the limit. We have
\begin{equation}
\rho_0 = \begin{cases} 
- \frac{2u}{\pi } & |v| \leq |u| \\
-\frac{1}{\pi } \left( \frac{u^2}{v} + v \right) & |v| > |u|.
\end{cases}
\end{equation}
If we look at the derivatives of the energy density, we see that the model exhibits a second order phase transition as we change $v$. The first derivative of $\rho_0$ is continuous, but there exists a discontinuity in the second derivative as
\begin{equation}
\frac{\partial^2 \rho_0}{\partial v^2}  = \begin{cases}
0 & |v| \leq |u| \\
- \frac{1}{\pi} \frac{u^2}{v^3} & |v| > |u|
\end{cases},
\end{equation}
revealing that the point $|v| = |u|$ corresponds to the critical point of a second order phase transition.

\begin{figure}[t]
\begin{center}
\includegraphics[width=0.6\columnwidth]{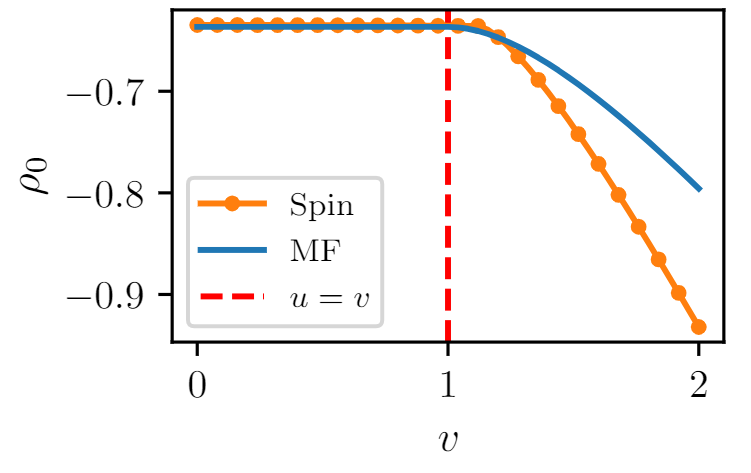}
\end{center}
\caption{A comparison of the ground state energy density vs. $v$ obtained from MPS simulation of the spin model from Ref.~\cite{Mathew} and the mean field (MF) approximation for $N= 200$ spins.}
\label{fig:E&chi_vs_v}
\end{figure}

In Fig.~\ref{fig:E&chi_vs_v}, we compare the ground state energy density vs. $v$ for the MPS numerics of the spin model \cite{Mathew} and the mean field approximation for a system of $N = 200$. We see that the mean field agrees well with the spin model, accurately predicting the location of the critical point. Below the critical point, the two models agree exactly, which suggests that the interactions induced by the chiral term are irrelevant in the ground state. Nevertheless, interactions become significant above the critical point.

\subsubsection{Central charge}

\begin{figure}[t]
\begin{center}
\includegraphics[width=0.475\columnwidth]{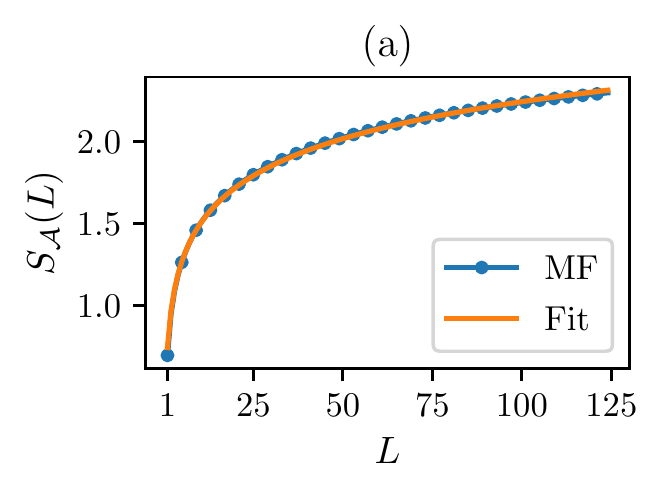}
\includegraphics[width=0.475\columnwidth]{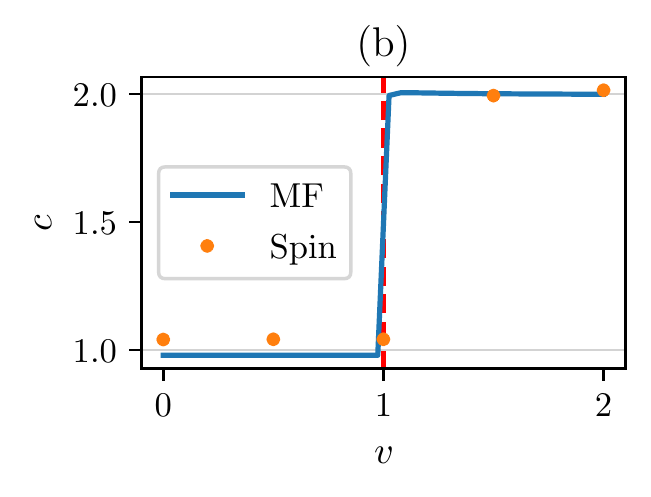}
\end{center}
\caption{(a) The entanglement entropy $S_L$ of the mean field (MF) model vs. $L$ for a system of size $N = 500$, for values $u = 1$ and $v = 0.5$ with periodic boundary conditions. We see the entanglement entropy follows Eq.~(\ref{eq:C4_entanglement_entropy}), allowing us to extract the central charge. (b) A comparison of the central charge $c$ of the mean field model and spin model vs. $v$ for the same system. We see that the central charge jumps from $c = 1$ to $c = 2$ across the phase transition for the mean field, suggesting that the degrees of freedom of the model have changed. The central charge for the spin model was calculated using OBC and fitted using Eq.~(2) in \cite{Entanglement}.}
\label{fig:C4_entanglement_entropy}
\end{figure}

To gain further insight into the nature of the chiral phase transition, we consider the behaviour of the ground state bipartite entanglement entropy as a function of $v$. Consider partitioning the system into two subsystems, $\mathcal{A}$ and $\mathcal{B}$, where $\mathcal{A}$ contains $L \ll N$ adjacent spins. We define the reduced density matrix of $\mathcal{A}$ as the partial trace over the remaining $N - L$ spins of $\mathcal{B}$ as $\rho_\mathcal{A} = \mathrm{Tr}_\mathcal{B}(\rho )$, where $\rho$ is the state of the whole system. As we are interested in the ground state only, we have $\rho = |\Omega \rangle \langle \Omega |$, where $|\Omega \rangle$ is the (pure) ground state of the total system. The entanglement entropy is defined as $S_\mathcal{A} = -\mathrm{Tr}(\rho_\mathcal{A} \ln \rho_\mathcal{A})$. As discussed above, the model is gapless for all $v$ so it can be described by a conformal field theory (CFT)~\cite{CFT}. In this case we expect the ground state entanglement entropy of a partition of  spins to obey the Cardy formula,
\begin{equation}
S^{PBC}_\mathcal{A}(L) = \frac{c}{3} \ln L + S_0, \label{eq:C4_entanglement_entropy}
\end{equation}
provided $L\ll N$, and where $c$ is the central charge of the CFT and $S_0$ is a constant \cite{Entanglement,Pasquale}, which applies to both the original spin model and the mean field approximation but differ for open (OBC) and periodic (PBC) and boundary conditions. We can measure the entanglement entropy of the mean field model (for PBC) quite simply by using the correlation matrix. We find that scaling behaviour of the entanglement entropy follows the formula in Eq. \ref{eq:C4_entanglement_entropy}, as shown in Fig.~\ref{fig:C4_entanglement_entropy}(a), allowing us to extract the central charge $c$ for various values of $v$. We find good agreement with the Cardy formula at even small $L$, we expect this is due to the primary operators that define the CFT being localised on single lattice sites, similar to the $XY$ model. Therefore no RG flow is necessary to access the universal behaviour.

Using the MPS results with open boundary conditions we compare the spin model and the mean field approximation. In Fig.~\ref{fig:C4_entanglement_entropy}(b) we see that $ c \approx 1 $ in the XX phase which jumps to $ c \approx 2$ in the chiral phase, with good agreement between the spin and mean field results. We can clearly interpret this in the mean field model: the additional Fermi points appearing when $|v| > |u|$ cause the model to transition from a $c=1$ CFT with a single Dirac fermion to a $c=2=1+1$ CFT with two Dirac fermions, as seen by the additional Fermi points of the dispersion in Fig.~\ref{fig:C4_lattice}(b). This can also be understood from the lattice structure of the MF model, as seen in Fig.~\ref{fig:C4_lattice}(a), where for $|v|\ll |u|$ a single zig-zag fermionic chain dominates ($c=1$) while for $|v|\gg |u|$ two fermionic chains dominate, corresponding to the edges of the ladder, thus effectively doubling the degrees of freedom ($c=2$).


\section{Emergent Black Hole Background}
\subsection{Diatomic model}

\begin{figure}
\includegraphics[width=\columnwidth]{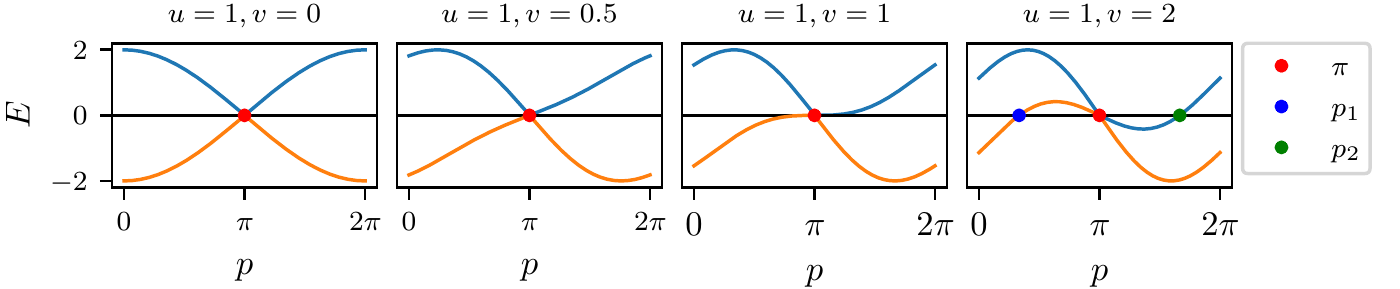}
\caption{The tilting of the Dirac cones as $v$ increases and $u = 1$. The blue and orange sections show the dispersions of the two operators $a$ and $b$ respectively in the diatomic unit cell for $a_c = 1$.}
\label{fig:C4_bh_disp}
\end{figure}

To make the link with relativity, the lattice sites are now labelled as alternating between sub-lattices $A$ and $B$ by introducing a two-site unit cell, as shown in Fig.~\ref{fig:C4_bh_disp}. The mean field Hamiltonian of Eq.~(\ref{eq:mf_ham}) can be reparameterised as 
\begin{equation}
H_\mathrm{MF} =  \sum_n -ua^\dagger_n(b_n + b_{n-1}) - \frac{iv}{2} (a_n^\dagger a_{n+1} + b^\dagger_n b_{n+1}) + \text{H.c.}, \label{eq:lattice_ham}
\end{equation}
where the fermionic modes $a_n$ and $b_n$ belong to sublattice $A$ and $B$, respectively, of the unit cell located at site $n$. These modes obey the commutation relations $
\{ a_n,a^\dagger_m \}  = \{ b_n, b_m^\dagger\} = \delta_{nm}$, while all mixed anti-commutators vanish. The index $n$ now labels the unit cells. A Fourier transform is performed on the Fermions with the definition 
\begin{equation}
a_n = \frac{1}{\sqrt{N_c}} \sum_{p \in \mathrm{B.Z.}} e^{i pa_\mathrm{c}n } a_p,
\end{equation}
and similarly for $b_n$, where $N_c = N/2$ is the number of unit cells in the system, $a_c = 2a$ is the unit cell spacing for a given lattice spacing $a$, and $\mathrm{B.Z.} = [0,2 \pi/a_c)$ is the Brillouin zone. The Fourier transformed Hamiltonian becomes
\begin{equation}
H_\mathrm{MF} = \sum_{p \in \mathrm{B.Z.}} \chi^\dagger_p h(p) \chi_p, \quad h(p) = \begin{pmatrix} g(p)  & f(p) \\ f^*(p) & g(p)  \end{pmatrix},
\end{equation}
where the two-component spinor is defined as $\chi_p = (a_p, b_p)^\mathrm{T}$ and the functions are given by
\begin{equation}
f(p) = -u(1+e^{-ia_\mathrm{c}p}), \quad g(p) = v \sin(a_\mathrm{c}p).
\end{equation}
As usual, the dispersion relation is given by the eigenvalues of the single-particle Hamiltonian $h(p)$ which yields 
\begin{equation}
\begin{aligned}
E(p) & = g(p) \pm |f(p)| \\
& =  v \sin(a_cp) \pm u \sqrt{2 + 2 \cos(a_\mathrm{c} p) } \label{eq:app_dispersion_2}.
\end{aligned}
\end{equation}
Clearly, this energy spectrum is not the same as the one we derive from Eq. (\ref{eq:C4_spin_ham}), however, we are only interested in the ground state properties, which are seen throughout this paper, such as in Fig.~\ref{fig:C4_jumps}, to agree exactly for $v < u$, and diverge slightly for $v > u$.

In Fig.~\ref{fig:C4_bh_disp}, it is found that the parameter $v$ has the effect of tilting the cones as it increases.
The Fermi points $\{ p_i\}$, defined as the points for which $E(p_i) = 0$, are found at
\begin{equation}
p_0 =  \frac{\pi}{a_\mathrm{c}}, \quad p_\pm  = \pm \frac{1}{a_\mathrm{c}} \arccos \left( 1 -  \frac{2u^2}{v^2} \right) .
\end{equation}
The roots $p_\pm$ only exist if the argument of $\arccos$ is in the range $[-1,1]$ which implies $|v| \geq |u|$ for these to appear in the dispersion. Therefore, if $|v| \leq |u|$, the only Fermi point is located at $p_0 = \pi/a_\mathrm{c}$ which is where the Dirac cone is located, as shown in Fig.~\ref{fig:C4_bh_disp}. When the cone over-tilts, so when $|v| \geq |u|$, then the additional zero-energy crossings at $p_\pm$ appear which is due to the Nielsen-Ninomiya theorem which states that the number of left- and right-movers must be equal \cite{NIELSEN1981219}.

\subsection{Continuum limit}

The continuum limit is obtained by Taylor expanding the single-particle Hamiltonian $h(p)$ about the Fermi point $p_0$ to first order in momentum which yields
\begin{equation}
h(p_0 + p)  = u\sigma^y p -  v \mathbb{I} p \equiv e_a^{\ i} \alpha^a p_i ,
\end{equation}
where we have set $a_\mathrm{c}=1$, the coefficients are defined as $e^{\ x}_0 = - v,e^{\ x}_1 = u $ and the Dirac matrices $\alpha^0 = \mathbb{I},\alpha^1 = \sigma^y $. 
Therefore, the continuum limit Hamiltonian after an inverse Fourier transform to real space is given by
\begin{equation}
H  =  \int_\mathbb{R} \mathrm{d}x \chi^\dagger(x) \left( -ie_a^{\ i} \alpha^a \overset{\leftrightarrow}{\partial_i} \right) \chi(x), \label{eq:cont_ham}
\end{equation}
with $A\overset{\leftrightarrow}{\partial_\mu}B =\frac{1}{2} \left( A \partial_\mu B - (\partial_\mu A)B \right)$ and the Dirac $\alpha^a = (\mathbb{I},\sigma^y)$ and $\beta = \sigma^z$. Note that the position is now measured in terms of unit cells with two lattice sites rather than single sites.

Comparing this Hamiltonian to the general one of Dirac particles in curved space \cite{S_M_Morsink_1991}, the continuum limit of the lattice model can be interpreted as a curved space field theory with zweibein
\begin{equation}
e_a^{\ \mu} = \begin{pmatrix} 1 & -v \\ 0 & u \end{pmatrix}, \quad e^a_{\ \mu} = \begin{pmatrix} 1 & v/u \\ 0 & 1/u \end{pmatrix} \label{eq:tetrad}
\end{equation}
and Dirac gamma matrices $\gamma^0  = \sigma^z$ and $\gamma^1 = -i \sigma^x$ which obey the anti-commutation relations $\{ \gamma^a, \gamma^b \} = 2\eta^{ab}$, with $\eta^{ab} = \mathrm{diag}(1,-1)$. The zweibein corresponds to the metric $g_{\mu \nu} = e^a_{\ \mu}e^b_{\ \nu} \eta_{ab}$ which gives
\begin{equation}
\mathrm{d}s^2 =  \left( 1 -\frac{v^2}{u^2} \right) \mathrm{d}t^2 -  \frac{2v}{u^2} \mathrm{d}t \mathrm{d}x - \frac{1}{u^2}\mathrm{d}x^2  \label{eq:app_GP_metric}. 
\end{equation}
If the variables $u$ and $v$ are upgraded to slowly-varying functions of space, then the preceding calculation is still valid and the event horizon is located at the point $x_h$, where $|v(x_h)| = |u(x_h)|$. In the small region in which $v$ is a slowly-varying functions of $x$ so the coupling of different momentum modes will be small and can be ignored to a good approximation, leaving 
the diagonal terms $a^\dagger_p a_p$ only. This is quite standard to do, for example in the SSH
model where the continuum is described by a Dirac equation \cite{Sabsovich}. For coordinate dependent coefficients, this is the Gullstrand-Painlev\'e metric \cite{volovik} also know as the \textit{acoustic metric} which is the Schwarzschild metric of a $(1+1)$D black hole expressed in Gullstrand-Painlev\'e coordinates. This metric is referred to here as an \textit{internal metric} of the model as it depends upon the internal couplings of the Hamiltonian and not the physical geometry of the lattice. In addition, this is a fixed classical background metric and the quantum fields have no back-reaction on the metric. 

In order to bring the metric Eq.~(\ref{eq:app_GP_metric}) into standard form, a coordinate transformation defined as $(t,x) \mapsto (\tau,x)$ is used, where
\begin{equation}
\tau(t,x)  = t - \int_{x_0}^x \mathrm{d} z  \frac{v(z)}{u^2 - v^2(z)} ,
\end{equation}
that maps the metric to
\begin{equation}
\mathrm{d}s^2 =  \left( 1 - \frac{v^2(x)}{u^2(x)} \right) \mathrm{d}\tau^2 -  \frac{1}{u^2(x) \left( 1 - \frac{v^2(x)}{u^2(x)} \right)} \mathrm{d}x^2, 
\end{equation}
which takes the general form of the Schwarzschild metric. As we are interested in the physics of the horizon we consider only the behaviour of the the system for $x\approx x_h$ where $f(x)$ can be taken to have a linear profile that changes sign around $x=x_h$ and satisfies $f(x_h)=0$, this condition corresponding to $v(x_h)=u$ in the couplings of our model. In the following we take $u(x) = 1$ so it aligns with the standard Schwarzschild metric in natural units. Quite remarkably, if we look at the condensed matter model with space-dependent parameters $u(x)$ and $v(x)$, then the phase boundary between the regions for $|v| < u$ and $|v| > u$, as seen in Fig. 8, can be interpreted in terms of a field theory with a black hole metric, where the phase boundaries align precisely with the event horizons of this metric.


\section{Chirality of the Model}

In this section we investigate the spin-chirality operator from Eq.~(\ref{eq:C4_chiral_op}) in detail. From the previous section, we see that the parameter $v$ has the effect of tilting the Dirac cones in the mean field description, as shown in Fig.~\ref{fig:C4_bh_disp}. Referring back to the original Hamiltonian of Eq.~(\ref{eq:C4_spin_ham}), we conclude that the chirality of the system is responsible for this tilting. This tilting emulates the over-tilting of a Dirac cone near a black hole, therefore it is of interest to study this operator and find what it can show about the the black hole system, especially in the case of the spins inside of the black hole horizon, corresponding to the transition into a chiral phase in a homogeneous lattice.

\begin{figure}[t]
\includegraphics[width=\linewidth]{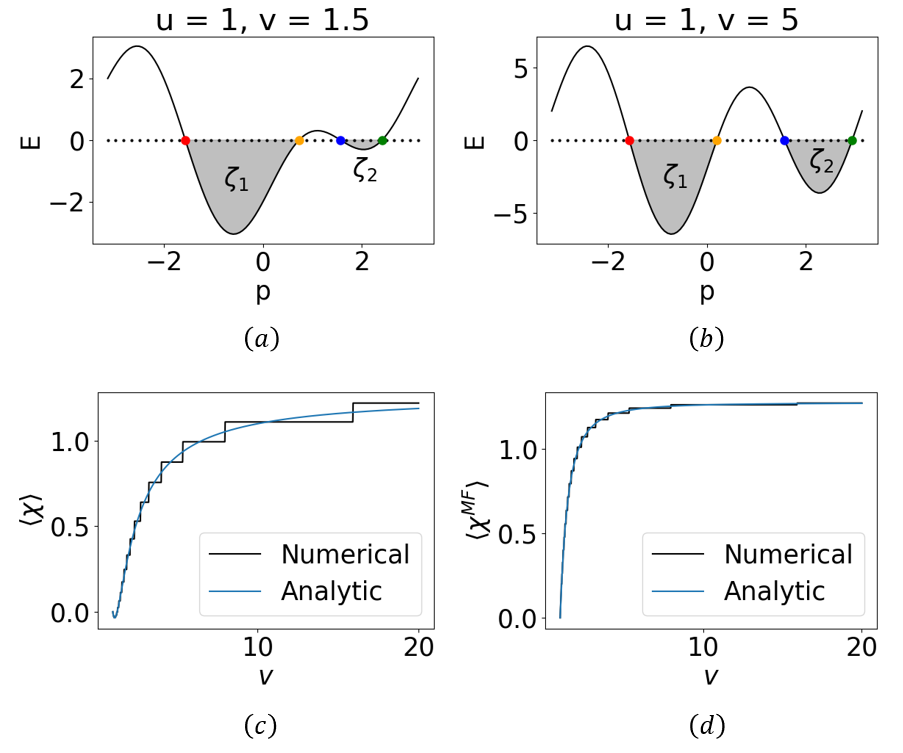}
\caption{The dispersion relations, with coloured Fermi-points corresponding to those in Fig.~(\ref{fig:C4_lattice})(b), when (a) $v = 1.5$, and (b) $v = 5$, showing the difference in number of discrete momentum states (equally space momentum from $-\pi$ to $\pi$), in the spaces $\zeta_1$ and $\zeta_2$. (c) The chirality of any single site in a 100 site homogeneous (`flat space') lattice as $v$ changes for the fully interacting chiral operator, with (d) giving the same for the mean field chiral operator. (c) and (d) also give the analytical solutions as given in Eq.~\ref{eq:C4_chi} and Eq.~\ref{eq:C4_MF_chi} respectively.}
\label{fig:C4_chiralities}
\end{figure}

Applying the Jordan-Wigner transformation to the chirality operator of Eq.~(\ref{eq:C4_chiral_op}), we arrive at the chirality operator in terms of fermionic modes given by
\begin{equation}
\begin{split}
\chi_n = & -2i (c^\dagger_n c_{n+1} + c^\dagger_{n+1} c_{n+2} - c^\dagger_n c_{n+2}) \\
        & +4i (c^\dagger_n c_{n+1} c^\dagger_{n+2} c_{n+2} + c^\dagger_{n+1} c_{n+2} c^\dagger_n c_n) + \mathrm{H.c.}.
        \end{split}
        \label{eq:full_chirality}
\end{equation}
It can be seen in Fig.~\ref{fig:C4_chiralities}(c) that the expectation of the chirality operator has a point after the transition $|v| > |u|$ where it is equal to $0$. If this operator is to be viewed as an order parameter, giving the transition at the point where $\chi$ is non-zero, it is unusual for it to return to this value. However, there is a choice now to be made of exactly which chiral operator to analyse. As the original spin Hamiltonian of Eq.~(\ref{eq:C4_spin_ham}) contains the chirality operator itself, applying MFT yields a non-interacting version of the chiral operator if we were to interpret this as the coefficient of $v/4$ in the mean field Hamiltonian of Eq.~(\ref{eq:mf_ham}). The MFT version of the chiral operator in fermionic form is therefore given by
\begin{equation}
\chi^\mathrm{MF}_n = 2ic^\dagger_n c_{n+2} + \mathrm{H.c.}, \label{eq:C4_chirality_MF}
\end{equation}
with ground state expectation given in Fig.~\ref{fig:C4_chiralities}(d). In the following we will consider both versions of the chirality operator as they give complementary information. We refer to Eq.~(\ref{eq:full_chirality}) as the full chirality, whilst Eq.~(\ref{eq:C4_chirality_MF}) as the mean field chirality.


\subsection{Discrete stepping of the chirality}

In Fig.~\ref{fig:C4_chiralities} it is shown how the chirality of a homogeneous chain system (corresponding to flat space) changes as the next-to-nearest neighbour terms in the Hamiltonian becomes more dominant. Clear discrete jumps in the value of chirality are found as the parameter $v$ is increased, corresponding to a momentum state leaving the left-hand Fermi sea (denoted as $\zeta_1$) as a momentum state enters the right-hand Fermi sea (denoted as $\zeta_2$). This is due to the fact that the number of discrete momentum states in the total Fermi sea is equal to $N/2$, i.e, $|\zeta_1| + |\zeta_2| = N/2$, where $|\zeta_i|$ is the number of momentum states in $\zeta_i$. As $v$ changes, these two disconnected regions of the Fermi sea change size and hence exchange states to keep the total fixed at $N/2$.

\begin{figure}[t]
\centering
  \includegraphics[width=0.8\linewidth]{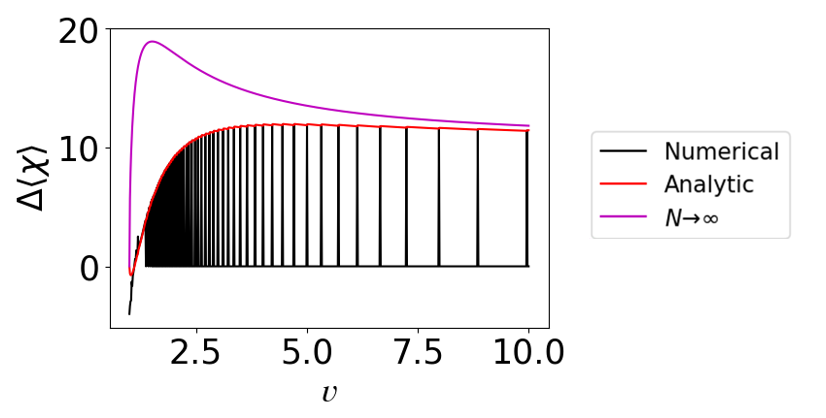}
  \caption{Measurements of the discrete jumping of total chirality as chiral coefficient $v$ is increased. The numerical solutions are found by taking the expectation of the chirality operator on the ground state at $v$, then subtracting that from the same operator for some small change $v + \epsilon$. The analytic solution is found from Eq.~(\ref{eq:jump_size}), and the thermodynamic solution $N \to \infty$ from Eq.~(\ref{eq:thermo_lim}).}
\label{fig:chi_jumps_measured}
\end{figure}

\begin{figure}[t]
\centering
  \includegraphics[width=\linewidth]{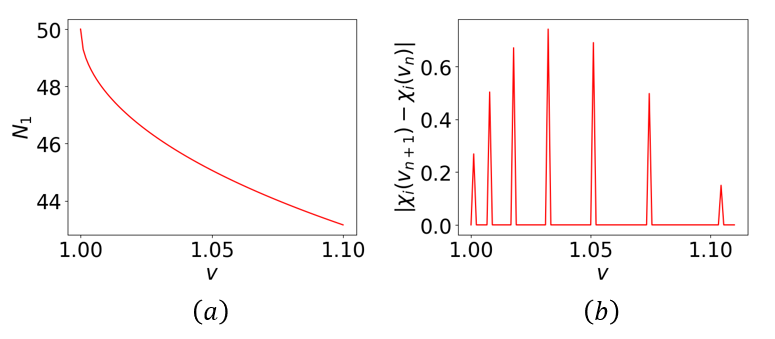}
  \caption{(a) The jumping rate of the chirality as found from Eq.~(\ref{eq:jump_freq}), (b) the discrete jumping found from applying Eq.~(\ref{eq:jump_size}). It is seen that as the number of states in $\zeta_1$ decreases, each integer change corresponds to a sudden increase in the chirality, with 7 total jumps in this interval.}
\label{fig:C4_jumps}
\end{figure}

We can determine analytically the behaviour of the chirality jumps shown in Fig.~\ref{fig:C4_jumps}. The total chirality in this instance can be diagonalised by the Fourier transform that also diagonalises the Hamiltonian of the system
\begin{equation}
c_n = \frac{1}{\sqrt{N}} \sum_{p \in \mathrm{B.Z.}} e^{ipan} c_p, 
\end{equation}
to give
\begin{equation}
\begin{split}
\langle\chi\rangle = & - 4\sum_{p \in \mathrm{B.Z.}} \sin(2ap) \\
& - \frac{8}{N}\sum_{p,k \in \mathrm{B.Z.}}\left[\sin(a(k-2p))+\sin(a(p-2k))\right],
\label{eq:jump_size}
\end{split}
\end{equation}
where the summed momenta $p$ and $k$ satisfy $E(p) \leq 0$ and $E(k) \leq 0$. The chirality is seen to jump in discrete steps as the chiral coupling $v$ is increased. This diagonalised total chirality can be used to find an analytical formula for the size of the jumps. We have
\begin{equation}
\begin{split}
&\lim_{\epsilon\to 0}\langle\chi(v + \epsilon)\rangle-\langle\chi(v)\rangle = \sqrt{1-\frac{u^2}{v^2}} \bigg[\frac{16u}{v} \\
& - \frac{32}{N}\sum_{p \in \mathrm{B.Z.}} \left(\sin(2p)-\frac{2u}{v}\cos(p)\right) \bigg].
\label{eq:jump_size_MF}
\end{split}
\end{equation}
In the large $v$ limit where the chiral operator becomes dominant we have
\begin{equation}
\lim_{v\to\infty}\Delta\langle\chi\rangle = -\frac{32}{N}\sum_p\sin(2p),
\end{equation}
which gives a value of about 10.2 for the limit the jumps tend towards as $v$ is increased. Additionally, we find that when $N$ becomes significantly large, then
\begin{equation}
\lim_{N\to\infty}\Delta\langle\chi\rangle = \sqrt{1-\frac{u^2}{v^2}} \left[\frac{16u}{v}+\frac{32}{\pi}\left(1+\frac{u^2}{v^2}\right) \right].
\label{eq:thermo_lim}
\end{equation}
Note that the order of the limits allows us to take $N \to \infty$ without forming a continuum version of the lattice model, as we already assumed the existence of the discrete stepping feature in Eq. (\ref{eq:jump_size_MF}). This analytically predicted behaviour of chirality jumps is in agreement with the numerical findings, as shown in Fig.~\ref{fig:chi_jumps_measured}.

The frequency of the jumps is controlled by the rate in which the momentum space covered by $\zeta_1$ shrinks while $\zeta_2$ grows, as shown in Fig.~\ref{fig:C4_chiralities}. This can be expressed in terms of the proportion of the momentum space in the Brillouin Zone that is spanned by $\zeta_1 = [\frac{-\pi}{2a}, p_1)$ for Fermi point $p_1 = \frac{1}{a}\sin^{-1}\left(\frac{u}{v}\right)$. The number of states in the left-hand Fermi sea is given by
\begin{equation}
N_1 = N \cdot \frac{p_1 + \frac{\pi}{2a}}{2\pi} = \frac{N\sin^{-1}(u/v) + N\pi/2}{2\pi a},
\label{eq:jump_freq}
\end{equation}
which, in Fig.~\ref{fig:C4_jumps}, shows a correspondence between $N_1$ dropping by an integer amount to the number of discrete steps in chirality taken.

\subsection{Chirality in the thermodynamic limit}

In this section we analyse the chirality in the thermodynamic limit for $N\rightarrow \infty$. Using the expression for the correlation matrix derived in Eqs.~(\ref{eq:C4_correlation_v<u}) and (\ref{eq:C4_correlation_v>u}), the ground state chirality with respect to the full chiral operator of Eq.~(\ref{eq:full_chirality}) is given by the simple expression
\begin{equation}
\begin{split}
\langle \chi_n \rangle & = 8\mathrm{Im}\big( C_{n,n+1}C_{n+2,n+2} - C_{n,n+2}C_{n+2,n+1} \\
& + C_{n+1,n+2}C_{n,n} - C_{n+1,n}C_{n,n+2} \\
&- ( C_{n,n+1} + C_{n+1,n+2} - C_{n,n+2} )/2\big) \\
& = \begin{cases}
0 & |v| < |u| \\
\frac{4}{\pi} \left(1 - \frac{u^2}{v^2}\right)\left(\frac{4u}{\pi v} - 1\right) & |v| \geq |u|
\label{eq:C4_chi}
\end{cases},
\end{split}
\end{equation}
and for the mean field chirality
\begin{equation}
\langle \chi^\mathrm{MF}_n \rangle = 4\mathrm{Im}\left( C_{n,n+2} \right) = 
\begin{cases}
0 & |v| < |u| \\
\frac{4}{\pi} \left(1 - \frac{u^2}{v^2}\right) & |v| \geq |u|
\end{cases}. \label{eq:C4_MF_chi}
\end{equation}
By Taylor expanding just above the critical point, we find the chirality goes as
\begin{equation}
\langle \chi_n \rangle \sim (v-v_\mathrm{c})^\gamma,
\end{equation}
where $v_\mathrm{c} = u$ is the critical point and $\gamma = 1$ is the critical exponent.
\begin{figure}[t]
\centering
  \includegraphics[width=0.6\linewidth]{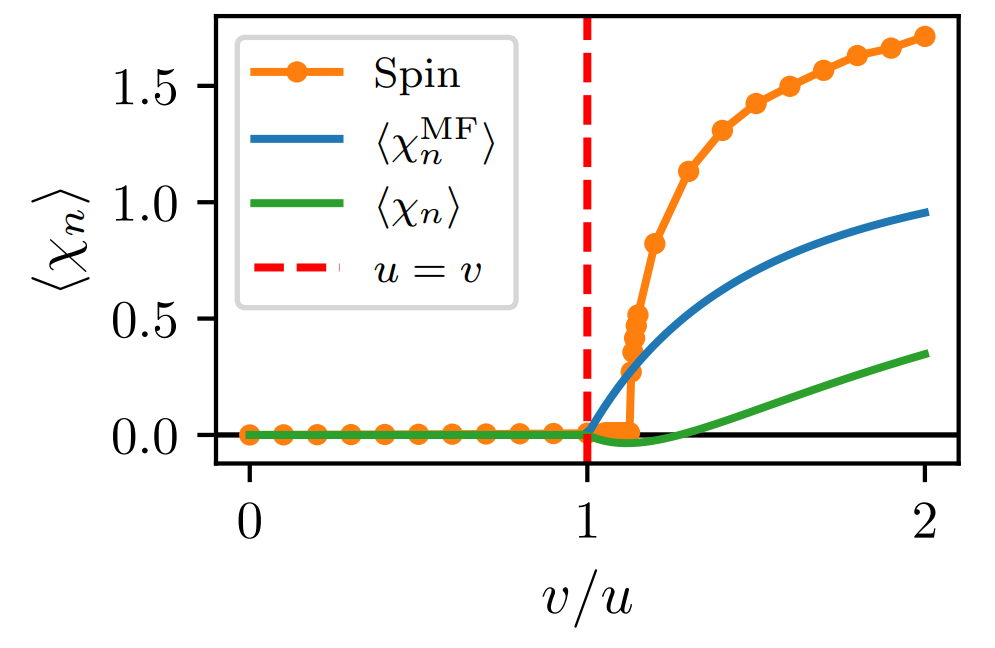}
  \caption{A comparison of the ground state chirality obtained from the mean field ground state $|\Omega\rangle$ using the two operators $\chi$ and $\chi^\mathrm{MF}$ (operators distinguished in Eq.~(\ref{eq:full_chirality}) and Eq.~(\ref{eq:C4_chirality_MF}) respectively), and the results obtained from exact diagonalisation of the spin model.}
\label{fig:C4_mps_comp}
\end{figure}
On the other hand, it was shown in Ref.~\cite{Mathew} by studying the full spin model of Eq.~(\ref{eq:C4_spin_ham}) using finite DMRG \cite{SCHOLLWOCK201196} that the phase transition of the full model is located at $v_\mathrm{c} \approx 1.12u$ with a critical exponent of  $\gamma \approx 0.39$. A comparison between the chirality of this MPS spin model simulation and the mean field approximation can be seen in Fig.~\ref{fig:C4_mps_comp}. The mean field faithfully captures the important information about the phase transition. In particular, just like for the energy density, the two models agree exactly below the critical point where the chirality is zero. The behaviour suggests the chirality is an order parameter for the model and emphasises again that, below the critical point, the interactions are irrelevant in the ground state. We see that the free fermion mean field approximation of Eq.~(\ref{eq:mf_ham}) accurately reveals that for small $v$ the system is in a disordered, gapless, XX phase, while as $v$ increases it passes through a second-order phase transition into a gapless chiral phase, corresponding to a non-zero ground state chirality $\langle \chi_n \rangle$.

From Eqs.~(\ref{eq:C4_chi}) and (\ref{eq:C4_MF_chi}), we see that the chirality is non-zero if and only if we have complex next-to-nearest-neighbour correlations $C_{n,n+2}$. We ask under what conditions is this the case. Consider a general tight-binding model with discrete translational symmetry and periodic boundary conditions. Suppose we had a model with inversion symmetry under the transformation $n \rightarrow  -n$ . This implies that the dispersion relation is an even function obeying $E(p) = E(-p)$, so our Fermi points come in $\pm$ pairs. Referring back to the definition of the correlation matrix in Eq.~(\ref{eq:C4_correlation_def}), we see that $C^*_{nm} = C_{nm}$ for an even dispersion relation: complex conjugation is equivalent to the transformation $p \rightarrow -p$ in the integral and, as the range of integral is symmetric under this transformation due to the Fermi points being $\pm$ symmetric as the dispersion relation is an even function, the integral, and hence correlation matrix, is invariant and hence real. In fact, this condition for inversion symmetry can be relaxed slightly: as long as the Fermi points come in $\pm$ pairs, even if the dispersion $E(p)$ itself is not an even function, the correlation matrix is real. This is the case for this model in the range $|v| \leq |u|$ as in this phase the Fermi points are fixed at $p_0 = \pm \pi/2a$ despite the dispersion itself not being even, as shown in Fig.~\ref{fig:C4_lattice}. However, this is broken when $|v| > |u|$ as new Fermi points appear and the correlation matrix is complex.


Let us now break inversion symmetry. A simple model that breaks inversion symmetry is a model with nearest-neighbour hoppings and complex couplings, with Hamiltonian
\begin{equation}
H = -ue^{-i \theta}\sum_n c^\dagger_n c_{n+1} + \text{H.c.},
\end{equation}
where $u \in \mathbb{R}$ and $\theta \in [0,2\pi)$. The breaking of inversion symmetry is apparent from the dispersion relation $E(p) = -2u \cos(p - \theta)$ as it is no longer an even function. The Fermi points of this model are at $p_0 = \theta \pm \pi/2$, therefore the correlations of this model are given by
\begin{equation}
\begin{aligned}
C_{nm} & = \frac{1}{2 \pi} \int_{\theta - \frac{\pi}{2}}^{\theta + \frac{\pi}{2}} \mathrm{d}p e^{-ip(n - m)} \\
& = \frac{\sin\left[(n-m)\frac{\pi}{2}\right]}{\pi(n-m)}e^{-i \theta(n-m)},
\end{aligned}
\end{equation}
which are complex, but notice that correlations between next-to-nearest-neighbours, where $|n -m| = 2$, are zero, therefore the chirality of this model will be zero too.

The simplest way to achieve complex next-to-nearest neighbour correlations is to include a term in the Hamiltonian which couples next-to-nearest neighbour sites and breaks inversion symmetry. A simple example of this is nothing but our mean field Hamiltonian of Eq.~(\ref{eq:mf_ham}). The interesting feature of this model is that for $|v| < |u|$, the dispersion relation retains its symmetric Fermi points at $p_\pm = \pm \pi /2$ despite the dispersion not being symmetric. Therefore, all correlators in this phase will be real as seen in Eq.~(\ref{eq:C4_correlation_v<u}) and hence the chirality will be zero. On the other hand, for $|v| > |u|$ the dispersion relation changes resulting in complex correlations which yields a non-zero chirality, giving the chirality its order parameter behaviour. 

\begin{figure}[t]
\begin{center}
\includegraphics[width=0.45\columnwidth]{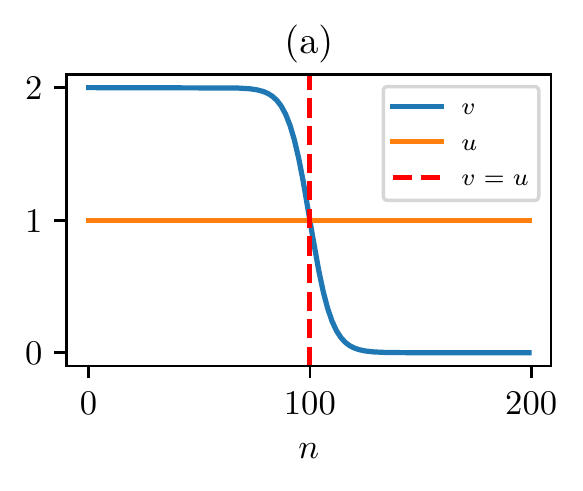}
\includegraphics[width=0.49\columnwidth]{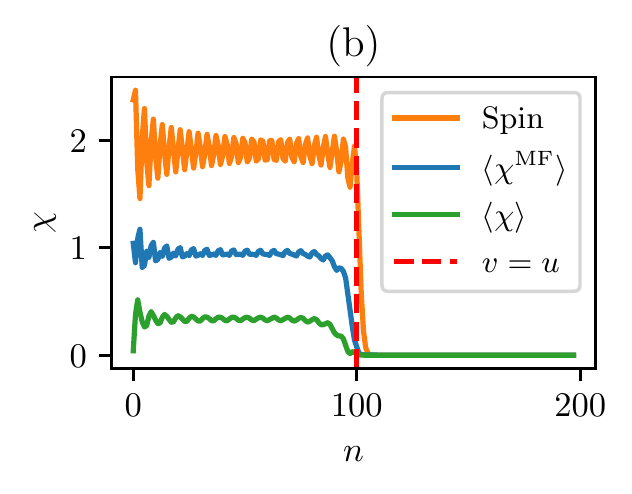}
\caption{(a) An example of an inhomogeneous distribution for the couplings $v$. (b) The corresponding chirality obtained from the spin model MPS \cite{Mathew} and the mean field model with the full operator $\chi$ . We see that the distribution of $v$ describes a phase boundary between a chiral ($v > u$) and non-chiral ($v <u$) phase.}
\label{fig:C4_inhomogenous_v}
\end{center}
\end{figure}

\subsection{Black hole profile chiralities}

The above analysis was conducted for homogeneous systems where $u$ and $v$ are constants. However, we still expect this to hold when we upgrade $v$ to a slowly varying function. We now consider profiles where $v(x)$ changes slowly and investigate the behaviour of the system around $v=u$, which from the emergent metric in Eq.~(\ref{eq:app_GP_metric}), corresponds to the location of the event horizon. In Fig.~\ref{fig:C4_inhomogenous_v} we present the chirality distribution across the system for a given coupling profile $v(x)$ for constant $u$. We observe the result that the system is  chiral where $|v| > |u|$, whereas for $|v| < |u|$ the system is non-chiral, therefore we have an interface between two phases.

\begin{figure}[t]
\centering
  \includegraphics[width=\linewidth]{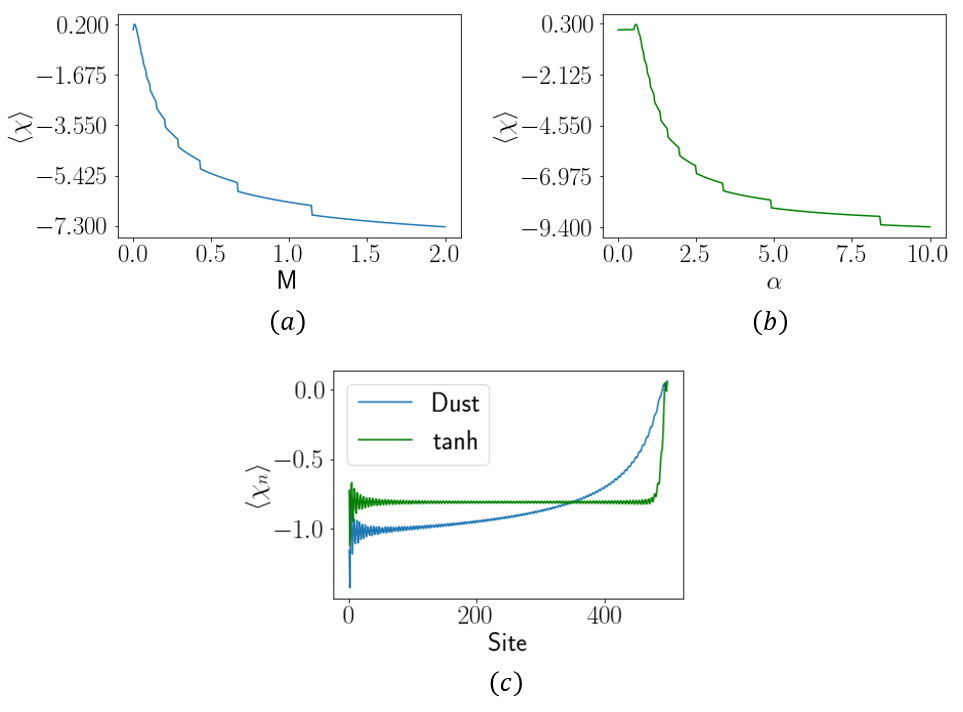}
  \caption{(a, b) Measurements of total chirality for 100 sites lattices with the horizon $x_h$ at site 95 for the collapsing dust profile when mass $M$ is increased, then, for the hyperbolic $\tanh$ profile as $\alpha$ is increased respectively. (c) Chirality expectations of each site of a 500 site lattice for spatial functions $v(x)$ with horizon $x_h$ positioned at site 450 for the collapsing dust metric with $M=1$ (blue) and with the hyperbolic profile of $v$ couplings as in Eq. (\ref{eq:tanh}) with $\alpha=2, \beta=0.2$ (green).}
\label{fig:C4_curv_chi}
\end{figure}

The chirality expectation can be found for different black hole backgrounds by choosing an appropriate relation for $v$. If a collapsing dust metric for a black hole is considered, the coupling becomes
\begin{equation}
v = \sqrt{1 - M(|x| - x_h/2)},
\end{equation}
where $M$ is the mass of the black hole and $x_h$ is the position of its horizon \cite{GR}. Another useful metric is a hyperbolic $\tanh$ profile \cite{shi2022onchip}, with coupling
\begin{equation}
v=\alpha[\tanh(\beta(x-x_h)+\delta)+1], \hspace{0.3cm} \delta=\tanh^{-1}\left(\frac{1}{\alpha}-1\right).
\label{eq:tanh}
\end{equation}

In Fig.~\ref{fig:C4_curv_chi} we present the chiralities across the lattice, where the position on the lattice $x$ corresponds to the position in space. Moreover, we present the total chiralities of these black hole profiles as their parameters are altered, giving similar results to those in the homogeneous case.

It has been shown that many analogue gravitational systems will exhibit a Hawking-like effect, whereby emission of radiation is described by scattering events following a thermal distribution at the Hawking temperature~\cite{Volovik1,Volovik2,Volovik3,Yang,Sabsovich,Maertens,BH_semimetal,Hang,Guan_2017,Retzker,Rodriguez,Roldan-Molina,Kosior,Steinhauer_2016,Stone_2013}. This has been explicitly demonstrated in the mean field model of Eq.~(\ref{eq:mf_ham}) with inhomogenous couplings, whereby interfaces between the two chiral phases thermalises the wavefunction~\cite{Mathew}. By preparing a single particle inside the horizon in the state $|n\rangle = c_n^\dagger|0\rangle$ and letting it evolve under the Hamiltonioan, it is found that once the wavefunction tunnels across the horizon the external distribution takes the form $P(k,t) \propto e^{-E(k)/T}$, where $T$ is the Hawking temperature of the effective black hole given by $T = |v'(x_h)|/2\pi$. This is an effective thermalisation exhibited shortly after the scattering~\cite{Yang,Wilczek}.


\section{Bosonization}

We now want to quantify the effect of interactions in our system introduced via the chirality term and analyse the validity of the mean field results. In higher dimensions, most interacting fermionic models can be studied using Fermi liquid theory. One dimensional systems can differ dramatically. The breakdown of Fermi liquid theory is intuitively explained by the nature of excitations near the Fermi surface \cite{Giamarchi}. In one dimension, the Fermi surface consists of two points, $k_F$ and $-k_F$. For inversion symmetric Hamiltonians, the dispersion in the vicinity of the Fermi surface is typically $\omega_1(k)=v_F(k-k_F)$ and $\omega_2(k)=-v_F(k-k_F)$. The \emph{nesting condition} $\omega_1(k)=-\omega_2(k)$ leads to a breakdown in perturbation theory and indicates that the interacting model differs dramatically from the noninteracting model. In fact, the low energy behaviour is typically described by collective, bosonic excitations using Luttinger liquid theory. 

For the model discussed here, bosonization has been previously employed when $|v| < |u|$ in which case there are only two Fermi points \cite{Mathew}. This resulted in the bosonized Luttinger liquid Hamiltonian
\begin{equation}
H = u \int dx \left[\Pi^2 + (\partial_x\Phi)^2 \right]
\end{equation}
for a bosonic field $\Phi$ with canonical momentum $\Pi$. This corresponds to a Luttinger coefficient \cite{Giamarchi} $K = 1$, corresponding to a free fermion model in the whole regime with the interactions simply renormalizing the Fermi velocities. This can be simply understood by noting that the Fermi velocities differ at the two Fermi points, therefore the nesting condition does not apply. 

When $|v| > |u|$ there are two additional Fermi points with equal Fermi velocites so the nesting condition becomes relevant. The bosonized system is now described by a four component Hamiltonian given, neglecting terms with minimal contribution, by
\begin{equation}
H = \sum_{\mu, \nu} \partial_x\phi_\mu h_{\mu\nu} \partial_x\phi_\nu
\end{equation}
where $\phi_\mu$ represents the bosonic fields \cite{miranda_2003} centred at each Fermi point in the sum of $\mu$ and $\nu$ and
\begin{equation}
h_{\mu \nu} = \frac{1}{\pi}\begin{pmatrix}\frac{\pi v_{L_1}}{2} - 2v & 0 & v-u & v-u\\ 0 & \frac{\pi v_{L_2}}{2} + 2v & u-v & u-v\\ v-u & u-v & \frac{\pi v_R}{2} + 2u & 2u\\ v-u & u-v & 2u & \frac{\pi v_R}{2} + 2u\end{pmatrix},
\end{equation}
with Fermi velocities $v_{L_{1, 2}} = 2(\mp u - v)$ and $v_R = 2v\left(1 - \frac{u^2}{v^2}\right)$. After a coordinate transformation we find, near the transition point $v \approx u$, the Hamiltonian takes the form
\begin{equation}
\begin{split}
H & = u \int dx \left[\Pi_1^2 + (\partial_x\Phi_1)^2 \right] \\
& + \sqrt{v_R v'_R} \int dx \left[\sqrt{\frac{v_R}{v_R'}}\Pi_2^2 + \sqrt{\frac{v_R'}{v_R}}(\partial_x\Phi_2)^2 \right],
\end{split}
\end{equation}
for bosonic fields $\Phi_1, \Phi_2$ and corresponding canonical momenta $\Pi_1, \Pi_2$, giving Luttinger coefficients of $K_1 = 1$ and $K_2 = \sqrt{v_R/v_R'} = \sqrt{v_R/(v_R + \frac{8u}{\pi})}$.
Since the Luttinger coefficient $K_2 \neq 1$, the interactions after the transition into the chiral phase are relevant and have a significant influence on the model.

\section{Classical Analysis}

\begin{figure}[t]
\centering
  \includegraphics[width=\linewidth]{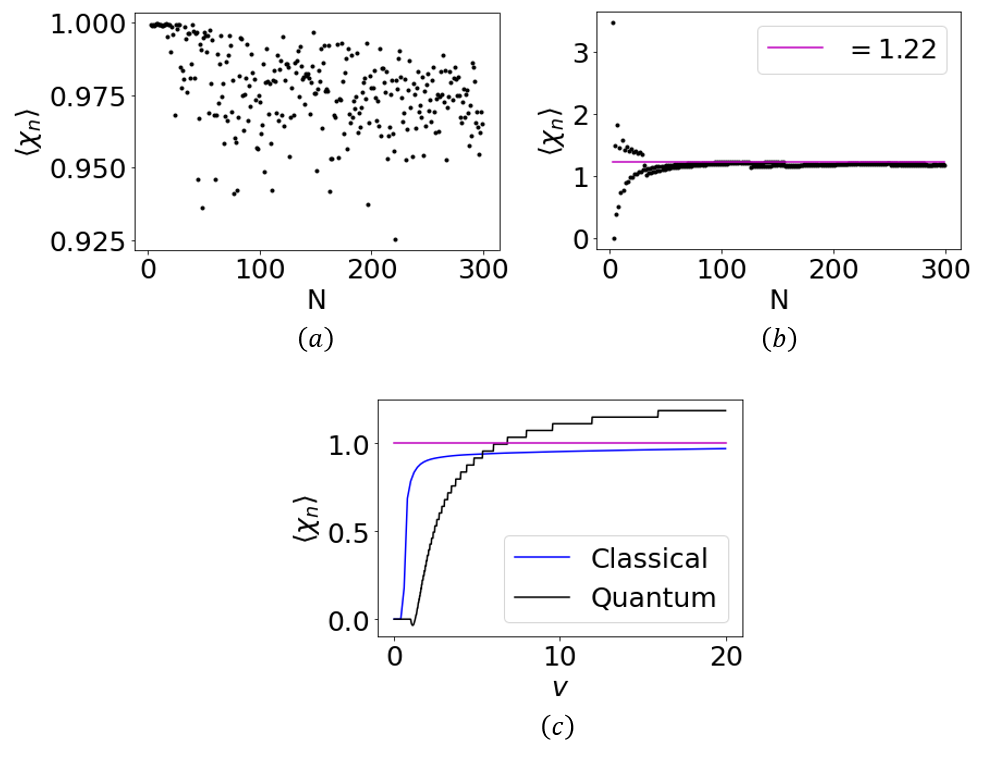}
  \caption{(a) Average chirality, when $u = 1$, $v = 20$, with the strong chiral contribution from large $v$ gives almost 1 as lattice size $N$ is increased for the classical chain, compared to the quantum case in (b) for values up to $N = 300$, which shows similar, locally oscillatory behaviour with a tendency toward $\sim 1.22$ as lattice size is increased. (c) Values for the classical/quantum chirality operator, given in blue/black, as $v$ is increased for a periodic lattice of $N = 300$. The classical value slowly approaches 1, whilst the quantum chirality grows larger towards $\sim 1.22$. It can also be seen that the classical chirality begins to grow before the transition point of $u = v$.}
\label{fig:C4_classic_chi}
\end{figure}

To gain further insight into the behaviour of the chiral spin chain we analyse the classical version of the model. Classically, the spins are unit vectors that can take arbitrary orientations. The dispersion is found by minimising the energy of the spin vectors in a classically equivalent energy function. The Hamiltonian is now given as
\begin{equation}
H=\sum_n\left[-\frac{u}{2}(S^x_nS^x_{n+1}+S^y_nS^y_{n+1})-\frac{v}{4}\vec{S}_n\cdot(\vec{S}_{n+1}\times\vec{S}_{n+2})\right],
\label{eq:C4_classic_e}
\end{equation}
for spin $\vec{S}_n=(\sin\phi_n\cos\theta_n,\sin\phi_n\sin\theta_n,\cos\phi_n)$. We adopt open boundary conditions, so the summation in Eq.~(\ref{eq:C4_classic_e}) ends at $N - 2$ for the chiral operator. The $u$ controlled XX portion of the energy tends to align all the spins with nearest neighbour couplings whilst the chiral coupling $v$ of the three spin interaction, tends to make neighbouring spins orthogonal. The overall spin configuration that minimises the energy was determined numerically, where the first site was set as spin up and all others sites were free, and the classical chirality found using DMRG.

\begin{figure}[t]
\centering
  \includegraphics[width=\linewidth]{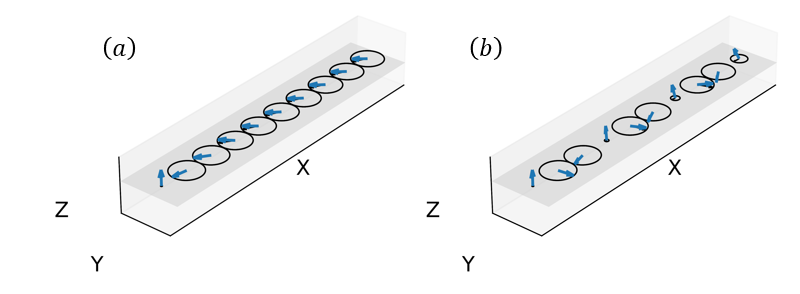}
  \caption{The procession of spins in a 10 spin lattice as found by the classical model for (a) $v = 0.8$ and (b) $v = 8$. In (b) the spin states are found in sets of 3 almost orthogonal spins that repeats and processes along the chain.}
\label{fig:spin_process}
\end{figure}

The average chirality has the form
\begin{equation}
\langle\chi\rangle = \frac{1}{N}\sum_n \vec{S}_n\cdot(\vec{S}_{n+1}\times\vec{S}_{n+2}).
\end{equation}
The value for chirality for the spin configurations that minimize the classical energy is given in Fig.~\ref{fig:C4_classic_chi}. By showing the spins along the chain in Bloch space it can be seen that, when $v > u$, the spin structure of the lattice is a repeating three spin sequence with these three spins almost orthogonal due to the chiral operator minimising whilst the spins are orthogonal, whereas the XX portion of the energy is minimised when the spins are parallel. This sequence then repeats along the lattice whilst slowly processing, as is shown in Fig.~\ref{fig:spin_process}, where the effect of increasing the chiral coupling strength is given.

The results from Fig.~\ref{fig:C4_classic_chi} show a similarity between the classical chirality calculations and those done in the quantum case. This figure also gives the changes in average chirality $\chi$ when the system size is increased, shown to be a tendency toward 1 in the classical regime as the spins align for every extension of the chain to maximise the chirality. From observing the spins Fig.~\ref{fig:spin_process} when $v$ is large it is found the spins take on a repeating 3-spin pattern in which they attempt to stay orthogonal to maximise $\chi$, which may increase by a maximum of 1 for every chiral operator acted along the chain. In contrast, in the quantum chain the chirality takes a maximal value of approximately 1.22, as shown in Fig.~\ref{fig:C4_classic_chi}. This indicates that chirality receives contributions from genuine quantum correlations, that cause its value to become larger than the maximum possible classical value of 1 \cite{Tsomokos_2008}.


\section{Conclusion}

While the 1-dimensional XX model supports the relativistic 1-dimensional Dirac equation, adding a chiral interaction causes the Dirac code to tilt an effect that is controlled by the chiral coupling. Surprisingly, this emulates the effect of gravitational background on Dirac fermions \cite{Mathew}. When the chiral coupling varies appropriately as a function of position then the chiral spin chain simulates the behaviour of Dirac particles in the black hole background.

In particular, we introduced a chiral spin model and simplified it with MFT in order to investigate its properties analytically. This included the dispersion relation, which exhibited transition shown by a splitting of the Fermi sea of the half filled ground state, and the central charge, giving what kind of CFT is defined by the transition of the system into a chiral phase. Results were then compared to an MPS simulation to give an idea of the accuracy of this mean field Hamiltonian.

Subsequently, we assessed the field theory defined by splitting the spin chain model into a diatomic unit cell, and found a comparison to the field theory of Dirac particles on the curved spacetime background of a black hole with the curvature determined by the couplings of the model. It is seen, by the dispersion relation of the diatomic model, that increasing the relative strength of the chiral coupling tilts the energy spectrum, analogous to the over-tilting of the Dirac cone as it enters the horizon of a black hole.

To determine the contribution of the chiral interactions in the behaviour of the system we employed the  bosonization method. This method determines the importance of the interactions, in turn giving an insight into the accuracy of the non-interacting approach that was taken in applying MFT to the Hamiltonian. Before the splitting of the Fermi sea, bosonization gave a Luttinger coefficient of $K = 1$, implying the interactions are insignificant. After the sea splits, i.e. for $v>u$, two separate coefficients emerge, corresponding to the two portions of the Fermi sea. One of them has a value different than 1, suggesting the interactions after the transition into the chiral phase have a significant influence on the model.

Finally, the classical version of the model is investigated. An energy function analogous to the quantum chain Hamiltonian was minimised in order assess the behaviour of chirality in the classical limit. We established that for large chiral coupling the spin vectors tend to be orthogonal, with a three spin pattern processing along the chain. Importantly, the quantum chain gives a value of chirality larger than then possible classical value thus demonstrating that quantum correlations contribute significantly in the behaviour of the system.
\newline\newline
We envision that our work can build the bridge between chiral systems and black holes, thus facilitating the quantum simulation of Hawking radiation, e.g. with cold atom technology. Moreover, our investigation opens the way for modelling certain strongly correlated systems by effective geometric theories with extreme curvature, thus providing an intuitive tool for their analytical investigation. As the bosonisation of the system in the chiral phase appears to indicate the interactions are important in this regime, a comparison between this model and that of a solveable quantum gravity could be a future focus of research, e.g. via measuring the scrambling of our model.

\section*{Author Contributions}

Ewan Forbes: conceptualization, methodology, investigation, software, writing - original draft. Matthew Horner: conceptualization, methodology, investigation, software, writing - review and editing. Andrew Hallam: methodology, investigation, software, data curation, writing - review and editing. Joseph Barker:  methodology, software, data curation, writing - review and editing. Jiannis Pachos: conceptualization, methodology, writing - review and editing.

\acknowledgements

We thank Patricio Salgado-Rebolledo for insightful discussions. E.F., M.D.H., A.H. and J.K.P. acknowledge support by EPSRC (Grant No. EP/R020612/1). JB acknowledges funding from a Royal Society University Research Fellowship. A. H. acknowledges
support by the Leverhulme Trust Research Leadership
Award RL-2019-015. Statement of compliance with EPSRC
policy framework on research data: This publication is
theoretical work that does not require supporting re-
search data.

\bibliography{citation}

\begin{appendix}

\section{On the space/time-likeness of the black hole}

In the Gullstrand-Painlev\'e frame, the metric is given by
\begin{equation}
ds^2 = \left( 1 - v^2 \right) dt^2 - 2 v dt dx -  dx^2
\end{equation}
where we have set $u = 1$. Consider the vectors $T^\mu = \delta^\mu_t$ and $X^\mu = \delta^\mu_x$ which point in the $t$- and $x$-directions respectively. We have
\begin{align}
g_{\mu \nu} T^\mu T^\nu & = g_{tt} = (1-v^2) \\
g_{\mu \nu} X^\mu X^\nu & = g_{xx} = - 1
\end{align}
so we see that the time coordinate is space-like inside the black hole ($v > 1$) and time-like outside ($v < 1$) the black hole as the norm of $T^\mu$ changes sign. However, the $x$-coordinate remains space-like everywhere as the norm of $X^\mu$ is negative everywhere. This is fine because, despite this, the metric is still Lorentzian everywhere. This means at every point there exists a space-like and time-like direction, they just do not coincide with the coordinates in the GP frame. To see this we just need to solve for the eigenvalues of the metric, which are given by
\begin{equation}
\mu_\pm = \frac{1}{2} \left( \pm \sqrt{v^4 + 4} - v^2 \right)
\end{equation}
where $\mu_+ > 0$ and $\mu_- < 0$ always. Therefore, the metric is Lorentzian and there will exist a pair of space-like and time-like directions at every point. This contrasts to a Euclidean metric which would have only positive eigenvalues.

If we change coordinates from GP coordinates to the Schwarzschild frame as $(t,x) \rightarrow (\tau,\chi)$ where
\begin{align}
\tau(t,x) & = t - \int_{x_0}^x dz \frac{v(z)}{1 - v^2(z)} \\
\chi(t,x) & = x
\end{align}
where we have explicitly used a different symbol for the spatial coordinate $\chi$ despite it being equal to $x$ always, then we have
\begin{equation}
ds^2 = (1-v^2)d\tau^2 - \frac{1}{(1-v^2)} d\chi^2 
\end{equation}
If apply the same procedure as before, we have
\begin{align}
g_{\tau \tau} & = 1-v^2 , \\
g_{\chi \chi} & = -\frac{1}{1-v^2}
\end{align}
which always have opposite signs, therefore if $\tau$ is time-like then $\chi$ is space-like, and vice versa. We see that in this coordinate system, the coordinate basis vectors do point along the space-like and time-like directions explicitly. Note that despite the transformation $\chi(t,x) = x$ being trivial, the basis vectors for $x$ and $\chi$ do not point in the same direction.

\section{Diagonalising the Chirality Operator}

The ground state expectation of the total chirality operator, given simply by the sum of the chirality for every site in the lattice, may be diagonalised by applying a the Fourier transform given in Eq.~(\ref{eq:C4_fourier_transform}). Utilising Wick's theorem \cite{Mathew}, the expectation of the chirality takes the form
\begin{equation}
\begin{split}
\langle\chi_i\rangle&=-2i(\langle c^\dagger_nc_{n+1}\rangle+\langle c^\dagger_{n+1}c_{n+2}\rangle+\langle c^\dagger_{n+2}c_n\rangle) \\
        &+4i(\langle c^\dagger_nc_{n+1}\rangle\langle c^\dagger_{n+2}c_{n+2}\rangle-\langle c^\dagger_nc_{n+2}\rangle\langle c^\dagger_{n+2}c_{n+1}\rangle \\
        &+\langle c^\dagger_{n+1}c_{n+2}\rangle\langle c^\dagger_nc_n\rangle-\langle c^\dagger_{n+1}c_n\rangle\langle c^\dagger_nc_{n+2}\rangle) + c.c.
\end{split}
\end{equation}
where $\langle c^\dagger_ic_j\rangle = \langle\Omega|c^\dagger_ic_j|\Omega\rangle$ where $|\Omega\rangle$ is the ground state of the system corresponding to half filling.The transform has the effect
\begin{widetext}
\begin{equation}
\begin{split}
\sum_n\langle\chi_n\rangle&=\sum_{p,q}-2i(e^{iaq}+e^{-ia(p-2q)}+e^{-2iap})(\frac{1}{N}\sum_ne^{-ian(p-q)})\langle c^\dagger_pc_q\rangle \\
&+\sum_{p,q,k,l}\frac{4i}{N}(e^{-ia(2k-q-2l)}-e^{-ia(2k-2q-l)}+e^{-ia(p-2q)} -e^{-ia(p-2l)})(\frac{1}{N}\sum_ne^{-ian(p-q+k-l)})\langle c^\dagger_pc_q\rangle\langle c^\dagger_kc_l\rangle+ c.c. \\
&=\sum_{p}-2i(2e^{iap}+e^{-2iap}) \langle c^\dagger_pc_p\rangle+\sum_{q,k,l}\frac{4i}{N}(2e^{iaq}-e^{-ia(k-2q)}-e^{-ia(q-2k)})\langle c^\dagger_{q+l-k}c_q\rangle\langle c^\dagger_kc_l\rangle+c.c. \\
&=\sum_{p}(8\sin(ap)-4\sin(2ap))\langle c^\dagger_pc_p\rangle +\frac{4i}{N}\sum_{p,k}\left(2e^{iap}-(e^{-ia(k-2p)}+e^{-ia(p-2k)})+c.c.\right)\langle c^\dagger_pc_p\rangle\langle c^\dagger_kc_k\rangle,
\end{split}
\end{equation}
\end{widetext}
which, considering the half filling ground state, simplifies to
\begin{equation}
\begin{split}
\sum_n\chi_n&=\sum_{p}(8\sin(ap)-4\sin(2ap))-\sum_p8\sin(ap) \\
&-\frac{8}{N}\sum_{p,k}\left(\sin(a(k-2p))+\sin(a(p-2k))\right) \\
&=-4\sum_{p}(\sin(2ap)) \\
&-\frac{8}{N}\sum_{p,k}(\sin(a(k-2p))+\sin(a(p-2k))),
\end{split}
\end{equation}
for the summations only accounting for momenta which produce negative energy solutions. Here, the fact that the ground state expectations follow the rule
\begin{equation}
\langle c^\dagger_pc_q\rangle =
\begin{cases}
\delta_{pq} & \text{if } E(p) \geq 0 \text{ and } E(q) \geq 0\\
0 & \text{if } E(p) < 0 \text{ or } E(q) < 0
\end{cases}
\end{equation}
meaning $\sum_p \langle c^\dagger_pc_p\rangle = \frac{N}{2}$ was utilised.

In order to find an analytical solution for the size and frequency of these jumps the fact the jumps originate from the 'hopping' of momentum states from the space $\zeta_1$ to the space $\zeta_2$ is utilised. If the chirality is known to change in a discrete step at certain points the parameter $v$ increasing by an infinitesimilly small amount $\delta v$, the amplitude in the difference in the total chirality is therefore given by $\chi(v_\delta) - \chi(v)$. From the Fermi points found in the dispersion relation Fig.~(\ref{fig:C4_lattice}) it is seen the momentum states that contributing to chirality differences as $v$ increases are those momenta lost at $v = \sin^{-1}(\frac{u}{v})$ and those gained at $v_\delta = \pi - \sin^{-1}(\frac{u}{v})$. This gives
\begin{equation}
\begin{split}
\chi(v_\delta)&-\chi(v)=4\left(\sum_{p_v}\sin(2ap_v)-\sum_{p_{v_\delta}}\sin(2ap_{v_\delta})\right) \\
&+\frac{8}{N}\sum_{p_v,k_v}(\sin(k_v-2p_v)+\sin(p_v-2k_v)) \\
&-\frac{8}{N}\sum_{p_{v_\delta},k_{v_\delta}}(\sin(k_{v_{\text{new}}}-2p_{v_\delta})+\sin(p_{v_\delta}-2k_{v_\delta})).
\end{split}
\end{equation}
Substituting in the identities for the momentum states gives
\begin{equation}
\begin{split}
\lim_{v_\delta\to v}\chi(v_\delta)&-\chi(v)=\frac{16u}{v}\sqrt{1-\frac{u^2}{v^2}} \\
&-\frac{32}{N}\sum_p\left(\sqrt{1-\frac{u^2}{v^2}}\sin(2p)-\frac{2u}{v}\sqrt{1-\frac{u^2}{v^2}}\cos(p)\right),
\end{split}
\end{equation}
which may be used to estimate the size of the jumps, as used in Fig.~(\ref{fig:C4_lattice}).

Different limits show that value of the jumps at different extremes. When the chiral term dominates $v \to \infty$, giving
\begin{equation}
\lim_{v\to\infty}\lim_{v_\delta\to v}\langle\chi(v_\delta)\rangle-\langle\chi(v)\rangle=\frac{-32}{N}\sum_p\sin(2p),
\end{equation}
which gives a value of about 10.2. In the thermodynamic limit $N \to \infty$ where $\frac{1}{N}\sum_p\rightarrow\frac{1}{\pi}\int_p$ the change in chirality becomes
\begin{equation}
\begin{split}
\lim_{N\to\infty}\lim_{v_+\to v}\langle\chi(v_+)\rangle-\langle\chi(v)\rangle&=\frac{16u}{v}\sqrt{1-\frac{u^2}{v^2}} \\
& +\frac{32}{\pi}\sqrt{1-\frac{u^2}{v^2}}\left(1+\frac{u^2}{v^2}\right)
\end{split}
\end{equation}
which is shown in Fig.~(\ref{fig:chi_jumps_measured}).

\section{Two Fermi-Point Bosonization}
\subsection{Normal Ordering}

The Hamiltonian may be separated into non-interacting and interacting systems $H = H_0 + H_{int}$ for
\begin{align}
H_0  & = \sum_n \left( - u  c_n^\dagger c_{n+1} - \frac{iv}{2} c^\dagger_n c_{n+2} \right) + \mathrm{H.c.}, \\
H_\mathrm{int} & = \frac{iv}{2} \sum_n  \left(c_n^\dagger c_{n+1} \sigma^z_{n+2} + c^\dagger_{n+1} c_{n+2} \sigma^z_n  \right)  + \text{H.c.} \label{eq:app_H_full_fermions}
\end{align}
Normal ordering, $: A : \ = A - \langle \Omega | A | \Omega \rangle$, affects Fermionic operators as
\begin{equation}
c^\dagger_n c_{n+1} = \ : c^\dagger_n c_{n+1} : + \langle \Omega| c^\dagger_n c_{n+1} |\Omega\rangle \equiv \ :c^\dagger_n c_{n+1} : + \alpha,
\end{equation}
and
\begin{equation}
\sigma^z_n = 1 - 2 c^\dagger_n c_n =  1 - 2( : c^\dagger_n c_n : + \langle \Omega | c^\dagger_n c_n| \Omega \rangle ) = -2 :c^\dagger_n c_n :,
\end{equation}
where the fact the ground state is a state of half filling gives $\langle \Omega|  c_n^\dagger c_n |\Omega \rangle = \frac{1}{2}$. Therefore, the interacting Hamiltonian becomes
\begin{equation}
\begin{aligned}
H_\mathrm{int} & = -iv \sum_n \left( :c^\dagger_n c_{n+1}: + \alpha \right) :c^\dagger_{n+2} c_{n+2}: \\
& + \left( :c^\dagger_{n+1} c_{n+2} :+ \alpha \right) :c^\dagger_n c_n: + \mathrm{H.c.} \\
& = -iv \sum_n :c^\dagger_n c_{n+1} : : c^\dagger_{n+2} c_{n+2} : \\
& + :c^\dagger_{n+1} c_{n+2} : : c^\dagger_n c_n : + \text{H.c.},
\end{aligned}
\end{equation}
where the fact that $\alpha$ is real and $:c^\dagger_n c_n:$ is Hermitian cancels $\alpha$ out of the calculation.

It has been found that in the non-chiral phase where $|v| < |u|$ there are two Fermi points at momenta $\pm\pi/2$.

\subsection{Field transformations}

 A real space continuum is approximated with field relations
\begin{equation}
\frac{c_n}{\sqrt{a}} = \sum_{\mu = \mathrm{R,L}} e^{ip_\mu an} \psi_\mu(x_n), \label{eq:expansion}
\end{equation}
where the sum is over the Fermi points, $\psi_\mu(x_n)$ is a slowly-varying continuous field sampled at discrete lattice sites $x_n = na$, with reinstated lattice spacing $a$.

First, we substitute the expansion of Eq.~(\ref{eq:expansion}) into $H_0$ of Eq.~(\ref{eq:app_H_full_fermions}) to give
\begin{equation}
\begin{split}
H_0  = & \sum_{\mu,\nu}\sum_n a e^{-i(p_\mu - p_\nu) an} -u  e^{ip_\nu a} \psi^\dagger_\mu(x_n)\psi_\nu(x_{n+1}) \\
& - \frac{iv}{2}  e^{2ip_\nu a} \psi^\dagger_\mu(x_n) \psi_\nu(x_{n+2}) + \text{H.c.}. \label{eq:H0_solved}
\end{split}
\end{equation}
We now discard any oscillating term in the Hamiltonian as these integrate to zero, so we requires $p_\mu = p_\nu$ in the first phase. This yields
\begin{equation}
\begin{aligned}
H_0 & = \sum_{\mu} \sum_n a \left( -u e^{ip_\mu a}  \psi^\dagger_\mu \psi_\mu + a \partial_x \psi_\mu \right) \\
& - \frac{iv}{2} e^{2ip_\mu a}  \psi^\dagger_\mu \left( \psi_\mu + 2a \partial_x \psi_\mu\right) +O(a^3)  + \text{H.c.} \\
& = -i \sum_{\mu} \sum_na^2 \left( \pm u \psi^\dagger_\mu \partial_x \psi_\mu  
- v \psi^\dagger_\mu \partial_x \psi_\mu \right) +O(a^3) + \text{H.c.}  \\
&  \rightarrow -2i\sum_{\mu} \int  \mathrm{d}x  v_\mu \psi_\mu^\dagger \partial_x \psi_\mu , 
\end{aligned}
\end{equation}
where in the third line $\pm$ corresponds to $\mu = \mathrm{R,L}$ for the right and left hand Fermi points, and we have renormalised the couplings as $au \rightarrow u$ and $av \rightarrow v$. We have defined 
\begin{equation}
v_\mathrm{R,L} = 2(\pm u - v), \label{eq:MF_fermi_v}
\end{equation}
which are the Fermi velocities $v_\mu = E'(p_\mu)$ obtained from the dispersion relation.

We now repeat the procedure for the interaction term $H_\mathrm{int}$ of Eq.~(\ref{eq:app_H_full_fermions}). We substitute in the expansion of Eq.~(\ref{eq:expansion}) into $H_\mathrm{int}$ to give
\begin{equation}
\begin{aligned}
H_\text{int} & = -iv   \sum_{\mu,\nu,\alpha,\beta} \sum_n ae^{-i(p_\mu - p_\nu + p_\alpha - p_\beta)an} \\
& \times (e^{i(p_\nu - 2(p_\alpha - p_\beta))a}+ e^{-i(p_\mu - 2p_\nu)a}) :\psi^\dagger_\mu \psi_\nu: : \psi^\dagger_\alpha \psi_\beta : \\
& +O(a^3) + \text{H.c.}, \label{eq:int}
\end{aligned}
\end{equation}
where we have expanded all fields to zeroth order in $a$ to ensure the Hamiltonian retains order $a^2$ and renormalised the couplings as $av \rightarrow v$. We discard any term that oscillates which requires $p_\mu - p_\nu + p_\alpha - p_\beta = 2n \pi/a$ for $n \in \mathbb{Z}$. With this we find only four terms survive giving us
\begin{equation}
\begin{split}
H_\text{int} & = 2v \int \mathrm{d}x \left( \rho_\mathrm{R}^2 + \rho_\mathrm{R} \rho_\mathrm{L}  - \rho_\mathrm{L} \rho_\mathrm{R} - \rho_\mathrm{L}^2 \right) + \text{H.c.} \\ & = 4v  \int \mathrm{d}x  \left( \rho^2_\mathrm{R} - \rho^2_\mathrm{L} \right),
\end{split}
\end{equation}
where we have defined the normal-ordered densities $\rho_{\mathrm{R,L}} = \ :\psi^\dagger_{\mathrm{R,L}} \psi_{\mathrm{R,L}}:$.

\subsection{Bosonising the Hamiltonian}

If we pull everything together, the normal-ordered Hamiltonian is given by
\begin{equation}
\begin{split}
:H: \ & = \ :H_0 + H_\text{int}: \ \\
& = -i\sum_{\mu = \mathrm{R,L}} \int  \mathrm{d}x  \left( v_\mu : \psi_\mu^\dagger \partial_x \psi_\mu :   \pm 4v  : \rho_\mu^2 : \right),
\end{split}
\end{equation}
where the $\pm$ corresponds to R and L respectively. Following \cite{miranda_2003}, we map the fermionic fields $\psi_\mu$ to bosonic fields $\phi_\mu$ with the mapping
\begin{equation}
\begin{split}
&\psi_\mathrm{R,L} = F_\mathrm{R,L} \frac{1}{\sqrt{2\pi \alpha}} e^{-i \sqrt{2\pi} \phi_\mathrm{R,L}} \\
& \rho_\mathrm{R,L} = \mp \frac{1}{\sqrt{2\pi}} \partial_x \phi_\mathrm{R,L},
\end{split}
\end{equation}
where $F_\mathrm{R,L}$ are a pair of Klein factors and $\alpha$ is a cut-off. The bosonic fields obey the commutation relations
\begin{equation}
[ \phi_\mathrm{R,L}(x),\phi_\mathrm{R,L}(y)] = \pm\frac{i}{2} \mathrm{sgn}(x-y),
\end{equation} 
whilst pairs of fields about different Fermi points commute. The fermionic fields and densities obey the identities
\begin{equation}
:\psi^\dagger_\mathrm{R,L} \partial_x \psi_\mathrm{R,L}: = \pm \frac{i}{2}  \partial_x \phi_\mathrm{R,L} , \quad \rho_\mathrm{R,L}  = \mp \frac{1}{\sqrt{2\pi}}\partial_x \phi_\mathrm{R,L},
\end{equation}
where we have taken lattice length $L \rightarrow \infty$. With this, the Hamiltonian is mapped to
\begin{equation}
\begin{aligned}
:H: \ &  =  \int \mathrm{d}x \frac{1}{2} \left[ |v_\mathrm{R}| : (\partial_x \phi_\mathrm{R})^2 : + |v_\mathrm{L}| :(\partial_x \phi_\mathrm{L})^2 : \right] \\
& + \frac{2v}{\pi} \left[ (\partial_x \phi_\mathrm{R})^2 - (\partial_x \phi_\mathrm{L})^2 \right] \\
& = \frac{1}{2} \int \mathrm{d}x \left( |v'_\mathrm{R}| :(\partial_x \phi_\mathrm{R})^2 :+ |v_\mathrm{L}'| : (\partial_x \phi_\mathrm{L})^2 : \right), \label{eq:h_boson}
\end{aligned}
\end{equation}
where the renormalised Fermi velocities are given by
\begin{equation}
v_\mathrm{R,L}' = 2 \left[ \pm u - v\left(1 - \frac{2}{\pi}\right) \right]. \label{eq:C4_renorm_fermi_velocities}
\end{equation}
As the Fermi velocities of the model are not equal, we must generalise the bosonization procedure of \cite{miranda_2003}. We define the canonical transformation
\begin{equation}
\begin{split}
&\Phi = \sqrt{\frac{\mathcal{N}}{2}} \left( \sqrt{|v_\mathrm{L}'|} \phi_\mathrm{L} - \sqrt{|v_\mathrm{R}'|} \phi_\mathrm{R} \right) \\
&\Theta = \sqrt{\frac{\mathcal{N}}{2}} \left( \sqrt{|v_\mathrm{L}'|}\phi_\mathrm{L} + \sqrt{|v_\mathrm{R}'|}\phi_\mathrm{R} \right), \label{eq:canonical}
\end{split}
\end{equation}
where $\mathcal{N}$ is a constant to ensure the fields obey the correct commutation relations. We require the fields $\Phi$ and $\Theta$ to obey the commutation relations \cite{miranda_2003}
\begin{equation}
[ \Phi(x), \Theta(y) ] = -\frac{i}{2} \mathrm{sgn}(x-y).
\end{equation}
In terms of our canonical transformation, we have
\begin{equation}
\begin{aligned} \relax
[\Phi(x),\Theta(y)] & = \frac{\mathcal{N}}{2} \left( |v'_\mathrm{L}|[\phi_\mathrm{L}(x),\phi_\mathrm{L}(y)] - |v_\mathrm{R}'| [\phi_\mathrm{R}(x),\phi_\mathrm{R}(y)] \right) \\
& = \frac{\mathcal{N}}{2}\left( -\frac{i|v_\mathrm{L}'|}{2} \mathrm{sgn}(x-y) - \frac{i|v_\mathrm{R}'|}{2} \mathrm{sgn}(x-y) \right) \\
& = - \frac{i\mathcal{N}}{4} (|v_\mathrm{L}'| + |v_\mathrm{R}'|) \mathrm{sgn}(x-y),
\end{aligned}
\end{equation}
therefore we require
\begin{equation}
\mathcal{N} = \frac{2}{|v'_\mathrm{L}| + |v_\mathrm{R}'|} = \frac{1}{2u}.
\end{equation}
Inverting the canonical transformation of Eq.~(\ref{eq:canonical}), we have
\begin{equation}
\sqrt{|v_\mathrm{L}'|}\phi_- = \sqrt{u} \left( \Theta + \Phi \right) , \quad \sqrt{|v_\mathrm{R}'|}\phi_+ = \sqrt{u} \left( \Theta - \Phi \right).
\end{equation}
Substituting this back into the bosonised Hamiltonian of Eq.~(\ref{eq:h_boson}) gives
\begin{equation}
:H: \ = u \int \mathrm{d}x \left[ :(\partial_x \Theta)^2: + :(\partial_x \Phi)^2: \right].
\end{equation}
Differentiating the commutator $[ \Phi(x),\Theta(y)]$ with respect to $y$, we find $[\Phi(x), \partial_y \Theta(y)] = i \delta(x-y)$, so we can identify the canonical momentum as $\Pi(x) = \partial_x \Theta(x)$. Therefore, the bosonised Hamiltonian takes the form of the free boson
\begin{equation}
:H: \ = u \int \mathrm{d}x \left[ :\Pi^2: + :(\partial_x \Phi)^2: \right],
\end{equation}
which is exactly the same result obtained from bosonising the XX model ($v = 0$). According to the theory of Luttinger liquids, this implies that $K = 1$ which is the sign of non-interacting fermions \cite{Giamarchi}, demonstrating that the interactions for $|v| < |u|$ are irrelevant in the ground state.

\section{Four Fermi-Point Bosonization}
\subsection{Bosonising for four Fermi points}

When $|v| > |u|$ the chirality kicks in and two extra points are formed, shown in Fig.~(\ref{fig:C4_lattice}). The Fermi points will here be labelled by where they appear on the two momentum spaces $\zeta_{1, 2}$, giving $p_{L_1,R_1} = \mp\pi/2,$ $p_{L_2} = \sin^{-1}(u/v),$ $ p_{R_2} = \pi-\sin^{-1}(u/v)$.

Just as before, the expansion of Eq.~(\ref{eq:expansion}) is substituted into $H_0$ of Eq.~(\ref{eq:app_H_full_fermions}) and oscillatory terms are disregarded; this time, for the four Fermi points. This again results in
\begin{equation}
H_0 = -i\sum_{\mu} \int  \mathrm{d}x  v_\mu \psi_\mu^\dagger \partial_x \psi_\mu, \label{eq:H0_four}
\end{equation}
where the couplings have been renormalised as $au \rightarrow u$ and $av \rightarrow v$. New couplings are defined as
\begin{equation}
v_\mathrm{L_2,L_1} = 2(\pm u - v), \quad v_\mathrm{R_1,R_2} = 2v\left(1 - \frac{u^2}{v^2}\right), \label{eq:MF_fermi_v_new}
\end{equation}
representing the four Fermi points where $v_{L_1}, v_{R_1} \in \zeta_1$ and $v_{L_2}, v_{R_2} \in \zeta_2$ with the $L$ and $R$ representing the Fermi points on the left and right sides of the $\zeta$ spaces.

Repeating the procedure for the interaction term $H_\mathrm{int}$, the new Fermi points are substituted into Eq.~(\ref{eq:int}). For the case of four Fermi points a more complicated contribution of
\begin{widetext}
\begin{equation}
\begin{split}
& H_\text{int} = av \sum_n \int \mathrm{d}x [ - 4(\rho_{L_1}\rho_{R_1} + \rho_{L_1}\rho_{R_2} + \rho_{L_2}\rho_{R_1} + \rho_{L_2}\rho_{R_2}) + 2Re(d)(\psi^\dagger_{R_1}\psi_{R_2}\psi^\dagger_{R_2}\psi_{R_1} + \psi^\dagger_{R_2}\psi_{R_1}\psi^\dagger_{R_1}\psi_{R_2}) \\
& + 2Re(a)(2\psi^\dagger_{L_1}\psi_{R_1}\psi^\dagger_{L_1}\psi_{R_2} + \psi^\dagger_{L_1}\psi_{R_1}\psi^\dagger_{R_1}\psi_{L_1} +\psi^\dagger_{L_1}\psi_{R_2}\psi^\dagger_{R_2}\psi_{L_1}
+\psi^\dagger_{R_1}\psi_{L_1}\psi^\dagger_{L_1}\psi_{R_1} + 2\psi^\dagger_{R_1}\psi_{L_1}\psi^\dagger_{R_2}\psi_{L_1} + \psi^\dagger_{R_2}\psi_{L_1}\psi^\dagger_{L_1}\psi_{R_2}) \\
&+ 2Re(c)(\psi^\dagger_{R_1}\psi_{L_2}\psi^\dagger_{L_2}\psi_{R_1} + 2\psi^\dagger_{R_1}\psi_{L_2}\psi^\dagger_{R_2}\psi_{L_2} + \psi^\dagger_{L_2}\psi_{R_1}\psi^\dagger_{R_1}\psi_{L_2}
+ 2\psi^\dagger_{L_2}\psi_{R_1}\psi^\dagger_{L_2}\psi_{R_2} + \psi^\dagger_{L_2}\psi_{R_2}\psi^\dagger_{R_2}\psi_{L_2} + \psi^\dagger_{R_2}\psi_{L_2}\psi^\dagger_{L_2}\psi_{R_2}) \\
& + 2Re(b)(\rho_{R_1}\rho_{L_1}
+ \rho_{R_1}\rho_{L_2} + \rho_{R_2}\rho_{L_1} + \rho_{R_2}\rho_{L_2} + 2\rho_{R_1}\rho_{R_2} + \rho^2_{R_1} + \rho^2_{R_2})],
\end{split}
\end{equation}
\vspace{-3em}
\end{widetext}
is obtained, with density functions as defined before, and factors given as
\begin{equation}
\begin{split}
Re(a) & = Re(e^{2isin^{-1}(u/v)} + ie^{-isin^{-1}(u/v)}) \\
& = 1 + \frac{u}{v} - \frac{2u^2}{v^2},
\end{split}
\end{equation}
\begin{equation}
Re(b) = Re(-2ie^{isin^{-1}(u/v)}) = \frac{2u}{v} ,
\end{equation}
\begin{equation}
\begin{split}
Re(c) & = Re(ie^{-isin^{-1}(u/v)} - e^{2isin^{-1}(u/v)}) \\
& = \frac{u}{v} - 1 + \frac{2u^2}{v^2},
\end{split}
\end{equation}
\begin{equation}
\begin{split}
Re(d) & = Re(ie^{3isin^{-1}(u/v)} - ie^{-3isin^{-1}(u/v)}) \\
& = \frac{8u^3}{v^3} - \frac{6u}{v}.
\end{split}
\end{equation}

\subsection{Bosonic fields}

Bosonic field equations are given as in the two Fermi point model
\begin{equation}
\psi_\mathrm{\mu} = F_\mathrm{\mu} \frac{1}{\sqrt{2\pi \alpha}} e^{\mp i \sqrt{2\pi} \phi_\mathrm{\mu}}, \quad \rho_\mathrm{\mu} = \mp \frac{1}{\sqrt{2\pi}} \partial_x \phi_\mathrm{\mu},
\end{equation}
where the $\mp$ results are for right movers $L_2, R_1, R_2$ (-) and left movers $L_1$ (+) (movement seen from their Fermi velocities Eq.~(\ref{eq:MF_fermi_v_new})). This gives the relation
\begin{equation}
:\psi^\dagger_\mathrm{R,L} \partial_x \psi_\mathrm{R,L}: = \pm \frac{i}{2}  \partial_x \phi_\mathrm{R,L},
\end{equation}
which allows the transformation $H_0$ from Eq.~(\ref{eq:H0_four}) into
\begin{equation}
H_0 = \frac{1}{2} \int \mathrm{d}x \sum_\mu |v_\mu|:(\partial_x\phi_\mu)^2:.
\end{equation}
The interacting Hamiltonian's transformation is

\begin{widetext}
\begin{equation}
\begin{aligned}
H_\text{int} &= \frac{2}{\pi} \int \mathrm{d}x [v(\partial_x\phi_{L_1}\partial_x\phi_{R_1} + \partial_x\phi_{L_1}\partial_x\phi_{R_2} - \partial_x\phi_{L_2}\partial_x\phi_{R_1} - \partial_x\phi_{L_2}\partial_x\phi_{R_2}) \\
& + \frac{v}{\pi} \left(\frac{u}{v} + 1 - \frac{2u^2}{v^2}\right) \left(\cos(\sqrt{2\pi}(2\phi_{L_1} + \phi_{R_1} + \phi_{R_2})) + 1\right) + \frac{v}{\pi} \left(\frac{u}{v} - 1 + \frac{2u^2}{v^2}\right) \left(\cos(\sqrt{2\pi}(2\phi_{L_2} - \phi_{R_1} - \phi_{R_2})) + 1\right) \\
& + u(\partial_x\phi_{R_1}\partial_x\phi_{L_2} + \partial_x\phi_{R_2}\partial_x\phi_{L_2} - \partial_x\phi_{R_1}\partial_x\phi_{L_1} - \partial_x\phi_{R_2}\partial_x\phi_{L_1} + 2\partial_x\phi_{R_1}\partial_x\phi_{R_2} + (\partial_x\phi_{R_1})^2 \\ 
& + (\partial_x\phi_{R_2})^2) + v((\partial_x\phi_{L_2})^2 - (\partial_x\phi_{L_1})^2) + \frac{4u^3}{v^2\pi} - \frac{3u}{\pi}],
\end{aligned}
\end{equation}
which, when ignoring the cosine terms, in matrix form, is given as
\begin{equation}
H = \left[\frac{1}{\pi}\begin{pmatrix}-2v & 0 & v-u & v-u\\ 0 & 2v & u-v & u-v\\ v-u & u-v & 2u & 2u\\ v-u & u-v & 2u & 2u\end{pmatrix} + \frac{1}{2}\begin{pmatrix} v_{L_1}& 0 & 0 & 0\\ 0 & v_{L_2} & 0 & 0\\ 0 & 0 & v_{R_1} & 0\\ 0 & 0 & 0 & v_{R_2}\end{pmatrix}\right] \begin{pmatrix} \partial_x\phi_{L_1} \\ \partial_x\phi_{L_2} \\ \partial_x\phi_{R_1} \\ \partial_x\phi_{R_2} \end{pmatrix}.
\label{eq:C4_matrix_H}
\end{equation}
\end{widetext}

\subsection{Luttinger coefficient}

It is possible to approximate to Eq.~(\ref{eq:C4_matrix_H})) as the cosine terms, if relevant in the renormalization group sense, would open a gap in the model \cite{miranda_2003}, meaning it may no longer be defined by a Luttinger liquid. However, it is seen from Fig.~\ref{fig:C4_entanglement_entropy}(b) that the chiral phase has a central charge $c = 2$, suggesting the model is 'doubly gapless', therefore, these terms cannot be significant and may be ignored.
In order to diagonalise the Hamiltonian Eq.~(\ref{eq:C4_matrix_H})), the coordinate transformations\\
\begin{equation}
\begin{split}
&\Phi_1 = \sqrt{\frac{1}{4u}}\left(\sqrt{|v'_{L_1}|} \phi_{L_1} - \sqrt{|v'_{L_2}|} \phi_{L_2}\right) \\
&\Theta_1 = \sqrt{\frac{1}{4u}}\left(\sqrt{|v'_{L_1}|} \phi_{L_1} + \sqrt{|v'_{L_2}|} 
\phi_{L_2}\right),
\end{split}
\end{equation}
\begin{equation}
\begin{split}
&\Phi_2 = \frac{1}{\sqrt{2}} \left(\varphi_{R_1} - \varphi_{R_2} \right) \\
&\Theta_2 = \frac{1}{\sqrt{2}} \left(\varphi_{R_1} + \varphi_{R_2} \right),
\end{split}
\end{equation}
are considered, which gives the Hamiltonian in the form 
\begin{equation}
H = \frac{1}{2}\text{diag}(u, u, v_R + 8u/\pi, v_R) + \frac{1}{2}\begin{pmatrix} 0 & M \\ M^\dagger & 0 \end{pmatrix}
\end{equation}
with $M = \frac{v-u}{\pi\sqrt{2u}} \begin{pmatrix} \sqrt{|v'_{L_1}|} - \sqrt{|v'_{L_2}|} & 0 \\ \sqrt{|v'_{L_1}|} + \sqrt{|v'_{L_2}|} & 0 \end{pmatrix}$. Therefore, close to the point of transition when $u \approx v$ this becomes
\begin{equation}
\begin{split}
H &= u \int dx \left[\Pi_1^2 + (\partial_x\Phi_1)^2 \right] \\
& + \sqrt{v_R v'_R} \int dx \left[\sqrt{\frac{v_R}{v_R'}}\Pi_2^2 + \sqrt{\frac{v_R'}{v_R}}(\partial_x\Phi_2)^2 \right].
\end{split}
\end{equation}
This gives a Luttinger coefficient of $K_2 = \sqrt{\frac{v_R}{v_R'}} = \sqrt{\frac{v_R}{v_R + \frac{8u}{\pi}}}$. The change in value of $K_2$ suggests that in the chiral phase, the model has significant interactions.

\end{appendix}

\end{document}